\documentclass[aps,prd,nofootinbib,twocolumn,superscriptaddress,preprintnumbers,balancelastpage,longbibliography,floatfix]{revtex4-1}
\usepackage{appendix}
\usepackage[utf8]{inputenc}
\usepackage{amsmath,amssymb,mathtools,bm}
\usepackage{graphicx, color, hepunits}
\usepackage[dvipsnames]{xcolor}
\usepackage{multirow}
\usepackage{tikz-feynman} 
 \usepackage{hyperref}
\hypersetup{
    colorlinks=true,       
    linkcolor=blue,        
    citecolor=blue,        
    filecolor=magenta,     
    urlcolor=blue          
}
\usepackage[english]{babel}

\usepackage{tensor}
\usepackage{soul}
\usepackage{booktabs}
\usepackage[capitalise, english]{cleveref}

\newcommand\blfootnote[1]{%
  \begingroup
  \renewcommand\thefootnote{}\footnote{#1}%
  \addtocounter{footnote}{-1}%
  \endgroup
}

\makeatletter
\providecommand*{\diff}%
	{\@ifnextchar^{\DIfF}{\DIfF^{}}}
\def\DIfF^#1{%
	\mathop{\mathrm{\mathstrut d}}%
		\nolimits^{#1}\gobblespace}
\def\gobblespace{%
	\futurelet\diffarg\opspace}
\def\opspace{%
	\let\DiffSpace\!%
	\ifx\diffarg(%
		\let\DiffSpace\relax
	\else
		\ifx\diffarg[%
			\let\DiffSpace\relax
		\else
  			\ifx\diffarg\{%
				\let\DiffSpace\relax
			\fi\fi\fi\DiffSpace}
   
\begin{document}

\title{A Pairwise Surrogate for Gravitational-Wave Spectra from Highly Relativistic Vacuum Bubble Collisions}

\author{Malte Buschmann}
\email{malte.buschmann@desy.de}
\affiliation{Deutsches Elektronen-Synchrotron DESY, Hamburg, Germany}

\author{Toby Opferkuch}
\email{toby.opferkuch@sissa.it}
\affiliation{SISSA International School for Advanced Studies, Via Bonomea 265, 34136, Trieste, Italy}
\affiliation{INFN, Sezione di Trieste, Via Bonomea 265, 34136, Trieste, Italy}

\preprint{SISSA 19/2026/FISI}
\date{\today}

\begin{abstract}
Gravitational waves from vacuum first-order phase transitions probe the nonlinear dynamics of relativistic bubble collisions.
Predicting their spectrum requires resolving Lorentz-contracted scalar bubble walls throughout a volume containing many bubbles, making the large-boost regime prohibitively expensive for direct lattice simulations.
We develop a simulation-based pairwise surrogate that separates the microscopic collision dynamics from the geometry of the multi-bubble transition.
This utilizes symmetry-reduced $(1+1)$-dimensional scalar-field simulations to build a library of two-bubble gravitational-wave spectra, including the nonlinear post-collision evolution.
These spectra are combined with geometric weights that account for screening by other bubbles.
We test the two-bubble spectra and the reconstructed spectra of three- and many-bubble systems against full $(3+1)$-dimensional lattice simulations.
Not only can the surrogate be orders of magnitude cheaper computationally, but it also provides access to highly relativistic collisions beyond the practical reach of direct many-bubble simulations.
This provides a route to studying how the spectral amplitude and shape depend on the collision boost and scalar potential.
\end{abstract}
\maketitle

\section{Introduction}
\label{sec:introduction}
\blfootnote{\newline The surrogate model is available on GitHub:\\\url{https://github.com/MSABuschmann/PhaseTransitionBuilder}}
Numerical studies of nonlinear cosmological dynamics are often controlled not by the largest or smallest physical scale separately, but by the hierarchy between them.
The simulation volume must contain the collective dynamics of interest, while the lattice spacing must resolve the localized structures that source it.
Axion-string evolution and radiation provide a longstanding example \cite{Davis:1986axions,Battye:1994radiation,Yamaguchi:1999strings}.
Simulations must follow a horizon-scale network while resolving microscopic string cores \cite{Fleury:2015strings,Gorghetto:2018strings,Buschmann:2021sdq}.

Vacuum first-order phase transitions provide a particularly severe example of this scale hierarchy \cite{Athron:2023review}.
This limit applies to sectors with negligible plasma interactions and approximates strongly supercooled transitions when friction and energy transfer to the plasma are subdominant \cite{Lewicki:2019collisions,Lewicki:2020azd}.
The metastable phase decays by nucleating bubbles of the stable vacuum \cite{Coleman:1977vacuum}, whose walls are accelerated by the released vacuum energy until collision \cite{Ellis:2020nnr,Gouttenoire:2021kjv}.
A cosmologically motivated Coleman--Weinberg benchmark with a model scale of $\SI{e+7}{\GeV}$ has an estimated collision boost $\gamma_*=\mathcal{O}(10^8)$, with almost all of the released energy in the bubble walls \cite{Levi:2022bzt}.
By contrast, the direct $(3+1)$-dimensional lattice simulations of Ref.~\cite{Cutting:2018tjt} considered nominal characteristic boosts only up to $\gamma_*\simeq 8$.

Bridging this gap with direct simulations is prohibitively expensive.
At large boost, the box size and physical evolution time grow as $\gamma_*$ at fixed potential and bubble number, while resolving the Lorentz-contracted walls requires the lattice spacing and time step to decrease as $\gamma_*^{-1}$.
For an evolution covering a fixed interval in units of the bubble separation, both the number of sites per spatial direction and the number of time steps therefore scale as $\gamma_*^2$, giving an evolution cost of $\mathcal{O}(\gamma_*^8)$.
These limitations have motivated both more efficient direct simulations and approximate descriptions of the gravitational-wave source. 
We review these approaches before introducing our pairwise surrogate, which combines the spectra from $(1+1)$-dimensional two-bubble simulations with a separate calculation of the multi-bubble geometry.

\begin{figure*}[t]
    \centering
    \includegraphics[width=\textwidth]{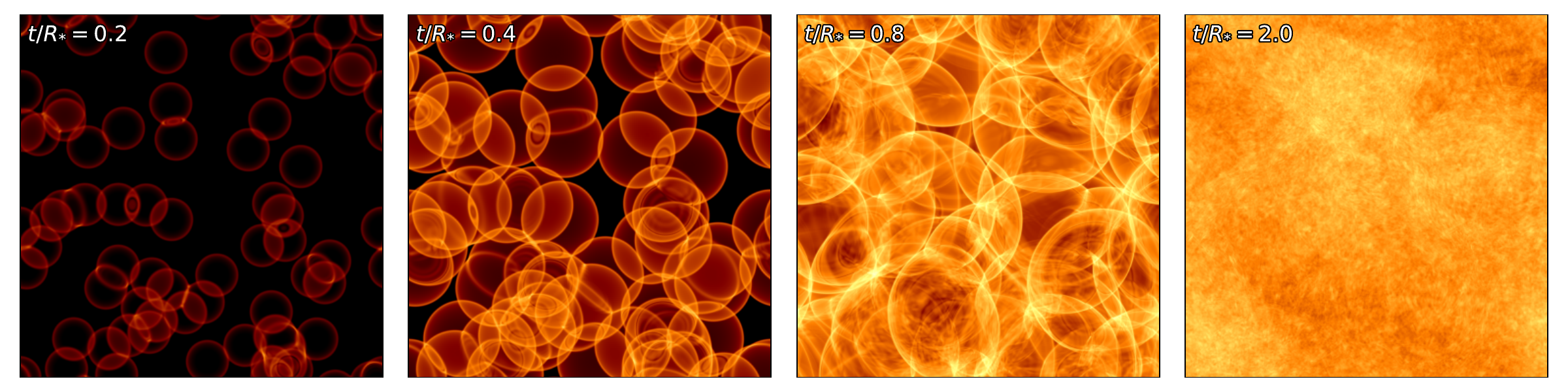}
    \caption{Snapshots of a vacuum first-order phase transition in a direct $(3+1)$-dimensional simulation with $N_b=64$ simultaneously nucleated bubbles and characteristic collision boost $\gamma_*\simeq4$.
    The panels show two-dimensional projections of the squared scalar-field time derivative $\dot\phi^2$ at $t/R_*=0.2$, $0.4$, $0.8$, and $2.0$, from left to right, where $R_*$ is the characteristic bubble separation.
    Field activity is initially concentrated near the expanding bubble walls, becomes more broadly distributed after collisions, and persists at late times. An animated version of this figure can be found \href{https://youtu.be/45Qiqm8tzOI}{online}.}
    \label{fig:bubbles}
\end{figure*}
 
A uniform lattice imposes the resolution needed for the contracted walls throughout the box, including regions where the field is nearly homogeneous.
Adaptive mesh refinement can therefore reduce the computational cost while the walls and collision regions remain well separated, using a fine grid near the walls and a coarser grid in most of the bubble interiors and surrounding false vacuum \cite{Buschmann:2024bfj}.
However, after collisions, outgoing wall remnants and the gradients left behind can require refinement across an increasing fraction of the simulation volume, as illustrated in \cref{fig:bubbles}.
If fine resolution is needed throughout most of the volume for a substantial part of the evolution, the computational advantage diminishes and the overhead of maintaining the adaptive mesh can render a fine uniform lattice more efficient.

Other approaches avoid this hierarchy altogether by replacing the scalar field with an effective gravitational-wave source.
The envelope approximation treats the walls as infinitesimally thin shells and removes each wall element after its first collision \cite{Kosowsky:1992rz,Huber:2008hg,Weir:2016tov}.
Its many-bubble spectrum can be computed either by following the uncollided surfaces numerically or by integrating the statistical two-point function of their stress \cite{Jinno:2016analytic,Megevand:2021features}.
The treatment of the walls' stress energy after collision is a separate physical assumption.
In the \emph{bulk-flow model}, thin shell elements continue along their original trajectories after the initial collision, with their energy diluted over an expanding surface \cite{Konstandin:2017bulk}.
More general thin-shell calculations allow a finite post-collision damping time \cite{Jinno:2017beyond}, while surface-source formalisms can accommodate non-spherical wall deformations \cite{Megevand:2021walls}.
These extensions demonstrate why geometric thinness alone does not imply the envelope result: Lorentz contraction does not specify how the anisotropic stress disappears or propagates after impact.
Direct scalar-field simulations find radiation from the subsequent field evolution, as well as spectral peaks and ultraviolet falloffs that differ from both the envelope and bulk-flow predictions \cite{Child:2012radiation,Cutting:2018tjt,Cutting:2020nla}.
Reduced-dimensional field calculations further show that the post-collision dynamics depend on the scalar potential \cite{Jinno:2019relativistic,Lewicki:2019collisions}, and that this dependence can remain visible in the gravitational-wave spectrum at substantial collision boosts \cite{Gould:2021dpm,Wang:2025thickness}.

A closer connection to the microscopic dynamics is provided by collision-calibrated thin-shell models.
Refs.~\cite{Lewicki:2020azd,Lewicki:2020jiv} use two-bubble simulations to determine the decay of the post-collision source, then impose this decay law in a many-bubble shell calculation.
Subsequent work incorporates the division of energy between scalar gradients and fluid motion \cite{Lewicki:2022sources} and extends both analytic envelope calculations and collision-calibrated shell models to expanding backgrounds \cite{Yamada:2025FLRW,Lewicki:2025expansion}.
These constructions already separate the local collision dynamics from the geometry of the transition.
They add the modelled shell amplitudes coherently, but encode the microscopic dynamics through a prescribed shell profile and decay law.

Like the collision-calibrated models above, we exploit the established $O(2,1)$ symmetry of an isolated vacuum-bubble pair to evolve the scalar field in $(1+1)$ dimensions \cite{Lewicki:2020azd,Lewicki:2020jiv,Hawking:1982collisions,Kosowsky:1992collisions,Gould:2021dpm}.
Rather than using this evolution to calibrate the decay of a thin-shell source, we calculate the two-bubble gravitational-wave spectra from the field profiles using numerical radiation integrals, avoiding the three-dimensional evolution of the six independent components of an auxiliary symmetric tensor.
The radiation calculation introduces a finite-duration cutoff, which breaks the ideal pair's boost symmetry; agreement with a lattice spectrum is therefore not guaranteed by the symmetry reduction of the scalar equations.
We test this prescription against full $(3+1)$-dimensional two-bubble simulations without applying the cutoff to the lattice source.
The resulting pair spectra form the microscopic input to our geometrically weighted many-bubble reconstruction.
Under the same assumptions, the symmetry-reduced $(1+1)$-dimensional scalar evolution instead scales as $\mathcal{O}(\gamma_*^4)$ per pair.
The numerical radiation integrals and the number of simulations required to build the pair library contribute to the costs of the surrogate; however, they are still vastly cheaper than any full $(3+1)$-dimensional lattice simulation and can be reused for any initial state realization.

In what follows, we consider identical bubbles nucleated at the same time.
We repeat the two-bubble calculation for different separations and potentials, recording the gravitational-wave spectrum as the collision evolves.
These spectra form a library that retains the nonlinear dynamics during impact and the subsequent relaxation of the scalar field.
We use this library to study how the amplitude and shape of the isolated-pair spectrum depend on collision boost and scalar potential, at boosts beyond the practical reach of direct many-bubble simulations.

For a specified configuration of bubbles, each pair separation selects the corresponding two-bubble spectrum from the library for the same scalar potential.
The influence of the remaining bubbles enters through a separate geometric calculation.
In the geometric calculation, the spherical wall surfaces intersect on a circle; we calculate the fraction of its circumference that lies outside all other bubble interiors.
This exposed fraction defines a time-dependent screening weight for the pair.
An additional bubble therefore changes the weight assigned to a pair, without requiring a new simulation of its field evolution.

We reconstruct the many-bubble spectrum by weighting the changes between successive finite-duration pair spectra, accumulating these changes over time and summing over all pairs.
Screening thus changes the weighting of subsequent spectral increments rather than deleting earlier contributions.
We consider both instantaneous turn-off of the screened fraction and a gradual, phenomenological decay informed by the relaxation of gradient energy in two-bubble simulations.
Separating this geometric calculation from the microscopic evolution allows the same collision library to be reused for different bubble configurations.
The library retains the nonlinear dynamics and interference within each isolated pair, but the reconstruction does not evolve the changes to those dynamics caused by additional collisions or include interference between distinct pairs.
Its accuracy therefore requires comparison with full many-bubble simulations.

The paper is structured as follows: \Cref{sec:setup} specifies the scalar theory, bubble initial data, and collision scales.
\Cref{sec:gw-methods} introduces the gravitational-wave observable and the finite-duration radiation prescription, then tests the latter against direct two-bubble simulations in \cref{sec:pair-comparison}.
\Cref{sec:surrogate-model} develops and tests the pairwise surrogate: \cref{sec:surrogate-construction} gives the geometric construction, \cref{sec:weight-decay} introduces relaxation after screening, and \cref{sec:three-bubble-closure,sec:many-bubble-validation} present the three- and many-bubble tests.
The large-boost applications are discussed in \cref{sec:ultrarelativistic-limit}.

\section{Scalar dynamics and collision scales}
\label{sec:setup}

\subsection{Phase-transition model}
\label{sec:model}

We consider a vacuum first-order phase transition in a single real scalar field.
The transition is assumed to complete on length and time scales much shorter than the Hubble scale, so that the scalar dynamics can be evolved in Minkowski spacetime and the gravitational radiation can be treated at linear order.
The scalar sector is described by
\begin{align}
    \mathcal{L} &= \frac{1}{2}\partial_\mu\phi\,\partial^\mu\phi-V(\phi)\,,
    \label{eq:lagrangian}\\
    V(\phi) &= \frac{1}{2}M^2\phi^2-\frac{\delta}{3}\phi^3
    +\frac{\lambda}{4}\phi^4\,.
    \label{eq:potential}
\end{align}
Here $M$ is the false-vacuum mass, with $V''(0)=M^2$, $\delta$ is the cubic coupling with dimensions of mass, and $\lambda$ is the dimensionless quartic coupling.
After fixing the overall mass and field normalization, the scalar dynamics depend only on the dimensionless combination \cite{Cutting:2020nla,Gould:2021dpm}
\begin{equation}
    \bar\lambda \equiv \frac{9\lambda M^2}{2\delta^2}\,.
    \label{eq:lambdabar}
\end{equation}
For $0<\bar\lambda<1$, the origin is a metastable false vacuum and the second minimum is the true vacuum.
The field values at the barrier top and true vacuum are
\begin{align}
    \phi_{\mathrm{top}}
    &=\frac{M}{\sqrt{8\lambda\bar\lambda}}
    \left(3-\sqrt{9-8\bar\lambda}\right)\,,\nonumber\\
    \phi_b
    &=\frac{M}{\sqrt{8\lambda\bar\lambda}}
    \left(3+\sqrt{9-8\bar\lambda}\right)\,.
    \label{eq:potential-extrema}
\end{align}
Defining $\Upsilon\equiv3+\sqrt{9-8\bar\lambda}$ and introducing the dimensionless quantities
\begin{equation}
    \bar x^\mu=Mx^\mu\,,\qquad
    \bar\phi=\frac{\phi}{\phi_b}\,,\qquad
    \bar V=\frac{V}{M^2\phi_b^2}\,,
    \label{eq:dimensionless-variables}
\end{equation}
gives
\begin{align}
    \frac{\mathcal{L}}{M^2\phi_b^2}
    &=\frac{1}{2}(\bar\partial\bar\phi)^2-\bar V(\bar\phi)\,,\\
    \bar V(\bar\phi)
    &=\frac{1}{2}\bar\phi^2
      -\frac{\Upsilon}{4\bar\lambda}\bar\phi^3
      +\frac{\Upsilon^2}{32\bar\lambda}\bar\phi^4\,.
    \label{eq:dimensionless-potential}
\end{align}
Thus the false and true vacua lie at $\bar\phi_f=0$ and $\bar\phi_b=1$.
Writing $\phi=\phi_b+\delta\phi$, the mass of linear scalar fluctuations about the true vacuum is $m_b^2=V''(\phi_b)$, or
\begin{equation}
    \frac{m_b^2}{M^2}=\bar V''(1)
    =\frac{\sqrt{9-8\bar\lambda}\,\Upsilon}{4\bar\lambda}\,.
    \label{eq:true-vacuum-mass}
\end{equation}
It sets the frequency of small, spatially homogeneous oscillations around that vacuum.
The numerical calculations use the rescaled variables in \cref{eq:dimensionless-variables}.
We retain bars when displaying rescaled quantities; otherwise, fields, coordinates and energies below are physical, with numerical lengths and times quoted in units of $M^{-1}$.
We focus on $\bar\lambda=0.845$\footnote{For the two- and three-bubble simulations we use $\bar\lambda=0.84$ while for the later multi-bubble results we use $\bar\lambda=0.845$.} and $\bar\lambda=0.069$, corresponding to relatively thin and thick initial walls.

\subsection{Bubble initial data and boost conventions}
\label{sec:bubble-scales}
We initialize the bubbles using the standard zero-temperature bounce construction \cite{Coleman:1977vacuum,Cutting:2020nla,Gould:2021dpm}.
We summarize the procedure here to specify the radius, wall-width and collision-boost conventions used to compare the reduced and direct simulations.
Let $\tau_E$ denote Euclidean time and $\boldsymbol x$ the spatial position relative to the bubble centre.
The $O(4)$-symmetric profile depends on the Euclidean radius $\rho_E=\sqrt{\tau_E^2+|\boldsymbol{x}|^2}$ and obeys
\begin{align}
    \frac{\mathrm{d}^2\phi_B}{\mathrm{d}\rho_E^2}
    +\frac{3}{\rho_E}\frac{\mathrm{d}\phi_B}{\mathrm{d}\rho_E}
    &=\frac{\mathrm{d}V}{\mathrm{d}\phi}(\phi_B)\,,\nonumber\\
    \phi_B'(0)&=0\,,\qquad \phi_B(\infty)=0\,.
    \label{eq:bounce}
\end{align}
We solve \cref{eq:bounce} using \textsc{CosmoTransitions} \cite{Wainwright:2011kj}.
Analytic continuation gives Lorentzian initial data with $\dot\phi=0$.
We define a reference radius and width from the initial field profile.
Writing $\phi_c\equiv\phi_B(0)$, the wall midpoint $r_{\mathrm{mid}}$ is the radius at which the field falls to half its central value, while $r_{\mathrm{in}}$ and $r_{\mathrm{out}}$ bracket the middle portion of the wall:
\begin{align}
    \phi_B(r_{\mathrm{in}})&=\frac{1+\tanh(1/2)}{2}\phi_c\,,
    &\phi_B(r_{\mathrm{mid}})&=\frac{1}{2}\phi_c\,,\nonumber\\
    \phi_B(r_{\mathrm{out}})&=\frac{1-\tanh(1/2)}{2}\phi_c\,,
    &\ell_0=r_{\mathrm{out}}&-r_{\mathrm{in}}\,.
    \label{eq:wall-contours}
\end{align}
The inner and outer radii correspond to approximately $73\%$ and $27\%$ of the central field value, and their separation defines the initial reference wall width $\ell_0$.
Using $\phi_c$ rather than the true-vacuum value $\phi_b$ also makes these definitions applicable to thick initial bubbles, whose centres need not yet have reached the true vacuum.
In the thin-wall limit, $\phi_c\simeq\phi_b$ and $r_{\mathrm{mid}}$ approaches the radius $R_c$ defined by $\phi_B(R_c)=\phi_b/2$, following the convention of Ref.~\cite{Cutting:2018tjt}.
Before interaction with another bubble, the surfaces at these fixed field values expand according to
\begin{equation}
    R_a(t)=\sqrt{r_a^2+(t-t_{\mathrm{nuc}})^2}\,,
    \quad a\in\{\mathrm{in},\mathrm{mid},\mathrm{out}\}\,.
    \label{eq:wall-trajectory}
\end{equation}
Here $t_{\mathrm{nuc}}$ is the nucleation time and $t\geq t_{\mathrm{nuc}}$; all configurations considered below use simultaneous nucleation at $t_{\mathrm{nuc}}=0$.
The decrease in the width between the inner and outer surfaces defines the reference contraction factor
\begin{equation}
    \gamma_{\mathrm{w}}(t)\equiv
    \frac{\ell_0}{R_{\mathrm{out}}(t)-R_{\mathrm{in}}(t)}\,.
    \label{eq:wall-boost}
\end{equation}
For two identical bubbles separated by a distance $d$, we define the reference collision time by the first meeting of their wall midpoints and evaluate this contraction factor at that time:
\begin{equation}
    d=2R_{\mathrm{mid}}(t_{\mathrm{coll}})\,,\qquad
    \gamma_{\mathrm{coll}}=\gamma_{\mathrm w}(t_{\mathrm{coll}})\,.
    \label{eq:midpoint-collision-convention}
\end{equation}
This assigns a collision-boost label $\gamma_{\mathrm{coll}}$ to each pair separation.
In the thin-wall limit, the contraction factor approaches the wall Lorentz factor.  
For a thick initial profile, $\gamma_{\mathrm{coll}}$ serves as a reference boost label and can differ from the Lorentz factor inferred from the motion of the evolved gradient-energy peak.
Geometric screening uses the measured radius of this peak, as described in \cref{sec:surrogate-construction}.

Throughout, we use ``thin-wall'' and ``thick-wall'' to distinguish the two parameter points according to their initial bubble profiles.
As the bubbles expand and accelerate, both develop Lorentz-contracted walls that become thin compared with the bubble radius.
We study the effects of thick walls on collision dynamics, together with ways of delaying this thinning, in a companion paper \cite{Buschmann:ThickBubbles}.

For a many-bubble simulation in a periodic box of side $L$ containing $N_b$ bubbles, the characteristic separation is
\begin{equation}
    R_*\equiv\frac{L}{N_b^{1/3}}\,.
    \label{eq:mean-separation}
\end{equation}
We characterize the collision scale of the simulation by $\gamma_*$, obtained by applying \cref{eq:midpoint-collision-convention} to a reference pair separated by $R_*$.
The boosts of individual collisions depend on the pair separations, so $\gamma_*$ characterizes the simulation volume rather than giving their average.

\section{Gravitational-wave calculations}
\label{sec:gw-methods}
\label{sec:numerical-overview}

We compute gravitational waves in two ways.
The direct calculation evolves the scalar field and tensor perturbations on a three-dimensional spatial lattice.
The reduced calculation evolves the scalar field in $(1+1)$ dimensions and obtains the radiation from numerical source integrals.
We first introduce both calculations, then explain why the reduced radiation source requires a finite-duration cutoff and test that prescription against full three-dimensional two-bubble simulations.

\subsection{Direct scalar and tensor evolution}

The direct calculation uses \textsc{Sledgehamr} \cite{Buschmann:2024bfj} to evolve the scalar equation on a three-dimensional spatial lattice,
\begin{equation}
    \ddot\phi-\nabla^2\phi+V'(\phi)=0.
    \label{eq:direct-scalar-evolution}
\end{equation}
The resulting scalar gradients source the tensor perturbations throughout the evolution.
At leading order in the gravitational coupling,
\begin{align}
    \ddot h^{\rm TT}_{ij}-\nabla^2h^{\rm TT}_{ij}
    &=16\pi G\,\Pi^{\rm TT}_{ij}\,,\nonumber\\
    \Pi^{\rm TT}_{ij}&=\Lambda_{ij,lm}
       (\partial_l\phi\,\partial_m\phi)\,.
    \label{eq:gw-evolution}
\end{align}
In Fourier space, $P_{ij}=\delta_{ij}-\hat k_i\hat k_j$ and $\Lambda_{ij,lm}=(P_{il}P_{jm}+P_{im}P_{jl}-P_{ij}P_{lm})/2$.
The scalar field is evolved without tensor backreaction, and the initial tensor perturbation and its time derivative vanish.

We use the full quadratic tensor energy,
\begin{equation}
    \rho_{\rm GW}(t)=\frac{1}{64\pi G}
    \left\langle
      \dot h^{\rm TT}_{ij}\dot h^{\rm TT}_{ij}
      +\partial_\ell h^{\rm TT}_{ij}\partial_\ell h^{\rm TT}_{ij}
    \right\rangle_V .
    \label{eq:gw-energy}
\end{equation}
Repeated spatial indices are summed, and the brackets denote a volume average.
Thus a Fourier mode contributes both $|\dot h^{\rm TT}_{ij}|^2$ and $k^2|h^{\rm TT}_{ij}|^2$.
The commonly used kinetic estimator, $\langle\dot h^{\rm TT}_{ij}\dot h^{\rm TT}_{ij}\rangle_V/(32\pi G)$, agrees with \cref{eq:gw-energy} after averaging freely propagating waves over their oscillations.
It need not agree at a particular snapshot, and equipartition cannot be assumed while a source remains active.

Our implementation is fully available on the \textsc{Sledgehamr} GitHub repository~\cite{sledgehamr_code}. We use a kick-drift-kick leapfrog algorithm to evolve the field in time with a 13-point stencil for the Laplacian, and a 5-point stencil for the gradients. The time step size is $\Delta t=0.2\Delta x$, and if not mentioned otherwise, our simulations use a lattice size of $2048^3$ cells. As we need to co-evolve seven scalar fields, we require 32 GPU-nodes of the Perlmutter supercomputer for a total of 128 Nvidia A100 GPUs. Even though \textsc{Sledgehamr} supports adaptive mesh refinement, we generally do not utilize this feature for the aforementioned reasons.  We tested a variety of higher-order integrators, such as the Runge--Kutta--Nyström method, as well as different stencil sizes and lattice spacings, and found that the above configuration agrees with our highest-resolution simulation to within $0.1\%$ around the peak and $\sim5\%$ in the UV tail. This configuration is similar to previous works such as that of~\cite{Cutting:2020nla}, albeit with a larger stencil size, as we found this to improve convergence noticeably. 

\subsection{Reduced scalar evolution and source integral}
\label{sec:two-bubble-setup}

For an isolated pair with symmetry-compatible initial data, boosts and rotations transverse to the collision axis leave the scalar solution invariant.
Taking that axis to be $z$, the field in the timelike region depends only on $z$ and $s=(t^2-x^2-y^2)^{1/2}$.
Its equation of motion becomes
\begin{equation}
    \partial_s^2\phi+\frac{2}{s}\partial_s\phi
    -\partial_z^2\phi+V'(\phi)=0\,.
    \label{eq:milne-eom}
\end{equation}
This $O(2,1)$ reduction retains the nonlinear collision dynamics without a thin-wall approximation \cite{Hawking:1982collisions,Kosowsky:1992collisions,Gould:2021dpm}.
For initial data that respect this symmetry, the reduced equation describes the same scalar solution as the full three-dimensional equation.
Approximate initial profiles, periodic images or additional bubbles can break the symmetry and invalidate this identification.

The solution supplies the scalar stress throughout the three-dimensional collision geometry.
The timelike region is obtained from the evolved $\phi(s,z)$, while the spacelike exterior is supplied by the continued isolated-bubble profiles.
The radiation integral weights this stress with the finite-duration cutoff $C_T(t)$ specified below.
We use the Fourier convention
\begin{align}
    S_{ij}(\boldsymbol k,t)
      &=\int\mathrm d^3x\,e^{-i\boldsymbol k\cdot\boldsymbol x}
          \partial_i\phi\,\partial_j\phi,\nonumber\\
    \mathcal T_{ij}(\boldsymbol k;T)
      &=\int_{t_i}^{t_f}\mathrm dt\,
          C_T(t)e^{ikt}S_{ij}(\boldsymbol k,t),
    \label{eq:source-fourier-integral}
\end{align}
where $k=|\boldsymbol k|$ and the time integral begins at the common nucleation time $t_i=0$.
Its upper endpoint $t_f$ is chosen beyond the source window, where the remaining contribution is negligible.
Analytic integration over the transverse azimuth reduces the integral to oscillatory kernels acting on the reduced field \cite{Kosowsky:1992collisions,Gould:2021dpm}.
The resulting spectrum is
\begin{equation}
    \frac{\mathrm dE_2^{\rm red}}{\mathrm d\ln k}
      =\frac{Gk^3}{2\pi^2}\int\mathrm d\Omega_{\hat{\boldsymbol k}}\,
          \left|\Lambda_{ij,lm}\mathcal T_{lm}\right|^2.
    \label{eq:reduced-radiation-spectrum}
\end{equation}
The norm sums the tensor indices, and the factors of $2\pi$ appropriate to our Fourier convention are included in the prefactor.
For an axisymmetric pair, the azimuthal integral is analytic.
The remaining integral over the polar angle $\theta$ relative to the collision axis uses 51 uniformly spaced values of $\cos\theta$ between zero and one, with the trapezoid rule and reflection symmetry accounting for both hemispheres.
These are angular quadrature points, not discrete Fourier modes of a spatial lattice.

The reduced calculation selects a finite source duration using a Gaussian turn-off,
\begin{equation}
    C_T(t)=
    \begin{cases}
       1,&t<T-t_0/2,\\
       \exp\!\left[-\ln2\left(1+\frac{2(t-T)}{t_0}\right)^2\right],
          &t\geq T-t_0/2,
    \end{cases}
    \label{eq:pair-window}
\end{equation}
with $t_0=0.03d$.
The source is unchanged until $T-t_0/2$, and the Gaussian reaches half height at $T$.
This is the Gaussian family used in Refs.~\cite{Kosowsky:1992collisions,Gould:2021dpm}, but our e-folding duration is $t_0/(2\sqrt{\ln2})\simeq0.018d$, shorter than the $0.03d$ chosen in Ref.~\cite{Gould:2021dpm}.
The numerical integration continues beyond $T$ and truncates the Gaussian only once its tail is negligible.
We denote this family of pair responses by
\begin{equation}
    \mathcal E^{\rm red}_2(k;d,T,C_T)
      \equiv\frac{\mathrm dE_2^{\rm red}}{\mathrm d\ln k}.
    \label{eq:pair-energy-definition}
\end{equation}

We solve Eq.~\eqref{eq:milne-eom} with a dedicated GPU code, in which both the field evolution and the source integral run in double precision on a single Nvidia A100 GPU. The field is evolved with a leapfrog scheme and a 3-point stencil for $\partial_z^2\phi$, with step size $\Delta s=0.2\,\Delta z$. The spacing $\Delta z$ is set by the width of the wall at the moment of collision, $\Delta z=(R_{\rm out}-R_{\rm in})/N_{\rm wall}$, with $N_{\rm wall}\geq50$, and the $z$ domain is padded such that it stays causally ahead of the latest time $T$ that is evaluated. The oscillatory kernels in the time integral of Eq.~\eqref{eq:source-fourier-integral} are integrated with the GPU-friendly Filon quadrature, using 40 panels per oscillation and at least 128 panels. These parameters have been tuned to ensure convergence. In particular, $N_{\rm wall}\geq50$ is required, as early collisions need to be simulated well beyond the time of initial point of contact, since those can take part in a phase transition for a very long time.  Since the field never leaves the GPU, the source integral is accumulated alongside the evolution, and because $C_T$ depends on $T$ only through a shift, a single run yields the spectra for a sequence of cutoff times $T$ at individual frequencies $k$. We describe this streaming implementation, which keeps the memory requirement independent of the length of the evolution and makes additional cutoff times inexpensive, in \cref{app:gpu}.

The physical role of the cutoff window and the comparison used to test it are discussed next.

\subsection{Finite-duration radiation and its validation}
\label{sec:observable-matching}

An exactly $O(2,1)$-invariant two-bubble spacetime cannot emit gravitational radiation \cite{Wu:1983gravitational,Kosowsky:1992collisions}.
This result assumes that the symmetry holds throughout the spacetime entering the radiation integral, without a finite-time boundary or source taper.
The scalar gradients can produce local anisotropic stress, but their contributions cancel in the outgoing radiation amplitude.
It does not follow that the tensor energy at a finite simulation time must vanish.
In particular, specifying finite-size bubbles on an initial time slice and evolving only their future is not the globally boost-invariant problem \cite{Kosowsky:1992collisions}.

A finite-duration prescription also breaks the boost symmetry.
Since transverse boosts mix Minkowski time with the transverse coordinates, a nonconstant $C_T(t)$ selects a frame and breaks the boost symmetry of the source entering \cref{eq:source-fourier-integral}.
It does not change the reduced scalar evolution.

Earlier two-bubble studies introduced source cutoffs to represent the finite duration of a collision in a transition containing other bubbles \cite{Kosowsky:1992collisions,Gould:2021dpm}.
The cutoff is therefore part of the radiation model, and its form is not fixed by the symmetry-reduced scalar equation.
Here we ask whether the prescription in \cref{eq:pair-window} reproduces the gravitational-wave spectrum of a two-bubble collision evolved directly in three spatial dimensions.

Our reference is the full tensor energy measured in \textsc{Sledgehamr} at time $T$, with no cutoff applied to its source.
The reduced calculation instead uses the Gaussian whose half-height time is $T$.
The two calculations therefore resolve different source histories even if their scalar solutions agree.
The direct measurement responds only to the source up to $T$, whereas the Gaussian suppresses some contributions before $T$ and retains a tail afterwards.
Their agreement is a numerical test of this prescription, not an identity between radiation extraction formulae.
The spectra and integrated energy histories in \cref{sec:pair-comparison} test it for both benchmark potentials and several collision boosts.

\subsection{Spectral normalization}

The normalization separates the overall gravitational strength from the shape of the collision spectrum.
Let $\rho_{\rm vac}=V(0)-V(\phi_b)$ be the released vacuum energy density and $\Omega_{\rm vac}=\rho_{\rm vac}/\rho_{\rm tot}$ its fraction of the total density at the transition.
For a fixed dimensionless source history, a stress of order $\rho_{\rm vac}$ acting over a length and time of order $R_*$ gives $h^{\rm TT}\sim G\rho_{\rm vac}R_*^2$ through \cref{eq:gw-evolution}.
The tensor energy then scales as $\rho_{\rm GW}\sim G\rho_{\rm vac}^2R_*^2$.
Dividing by $\rho_{\rm tot}$ and using $H_*^2=8\pi G\rho_{\rm tot}/3$ gives the overall scaling $\Omega_{\rm GW}\propto(H_*R_*\Omega_{\rm vac})^2$.
We remove this factor by plotting
\begin{equation}
    \mathcal P_{\rm GW}(k,t)=
    (H_*R_*\Omega_{\rm vac})^{-2}
    \frac{\mathrm d\Omega_{\rm GW}(k,t)}{\mathrm d\ln k},
    \label{eq:main-spectrum-normalization}
\end{equation}
Here $\Omega_{\rm GW}=\rho_{\rm GW}/\rho_{\rm tot}$ and $H_*$ is the Hubble rate at the transition.
Dependence on the potential, collision boost, and post-collision dynamics remains in $\mathcal P_{\rm GW}$.
Using $\mathrm dE_{\rm GW}=V\,\mathrm d\rho_{\rm GW}$ for the physical simulation volume $V$, the same normalization converts the reduced energy spectrum to
\begin{equation}
    \mathcal P_{\rm GW}
      =\frac{3}{8\pi G\rho_{\rm vac}^2R_*^2V}
          \frac{\mathrm dE_{\rm GW}}{\mathrm d\ln k}.
    \label{eq:absolute-energy-normalization}
\end{equation}
For many bubbles, $R_*$ is given by \cref{eq:mean-separation}.
The few-bubble plotting conventions are stated with their comparisons.

\subsection{Isolated two-bubble comparison}
\label{sec:pair-comparison}

\begin{figure*}[t]
    \centering
    \includegraphics[width=\textwidth]{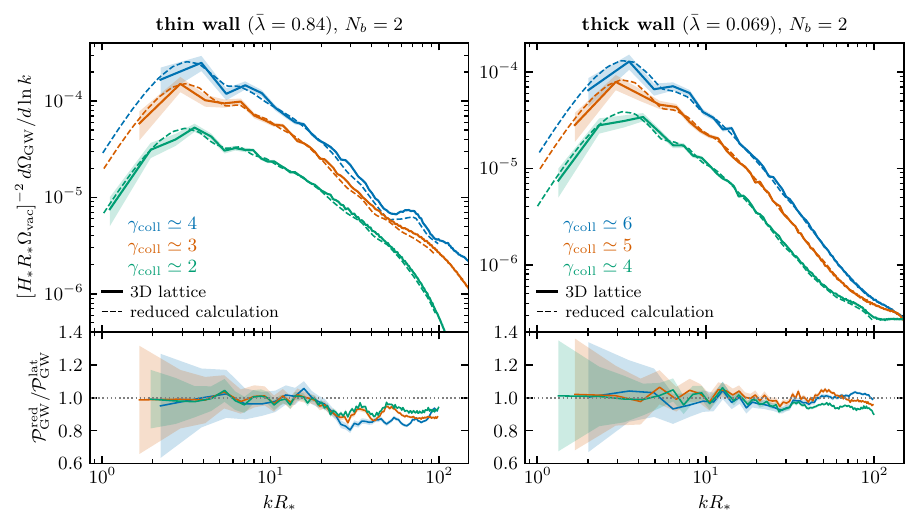}
    \caption{Gravitational-wave spectra from collisions of two isolated bubbles, comparing a reduced-dimensional calculation that exploits the collision symmetry with full $(3+1)$-dimensional lattice simulations.
    The left and right panels show $\bar\lambda=0.84$ and $0.069$, respectively.
    Blue, orange, and green identify $\gamma_{\rm coll}\simeq4,3,2$ in the left column and $6,5,4$ in the right column.
    Solid curves are the full lattice spectra, while dashed curves are the reduced two-bubble predictions with the smooth source cutoff of \cref{eq:pair-window}.
    We use the bubble separation $d$ as the reference length, so $R_*=d$.
    The lattice spectrum is taken from the saved snapshot closest to $t_m=1.095d$.
    For the reduced calculation, we choose the source window to fall to half its maximum at the time of that snapshot.
    The normalization is given in \cref{eq:main-spectrum-normalization}.
    The lower panels show $\mathcal P_{\rm GW}^{\rm red}/\mathcal P_{\rm GW}^{\rm lat}$ after applying the lattice's radial shell-sum estimator to the reduced response, including the discrete mode multiplicities.
    Only complete shells within the reduced frequency coverage are retained.
    Upper spectra retain their original sampling.
    Shading shows an approximate mode-counting estimate, not a numerical convergence band.
    This comparison tests the finite-duration radiation prescription against unwindowed tensor evolution, as discussed in \cref{sec:observable-matching}.}
    \label{fig:two-bubble-comparison}
\end{figure*}

Earlier comparisons have tested the symmetry-reduced scalar dynamics against (3+1)-dimensional evolution \cite{Bond:2015oscillons}.
Here we test whether the finite-duration reduced radiation prescription reproduces the absolute gravitational-wave spectra measured in unwindowed direct tensor evolution, as described in \cref{sec:observable-matching}.
We compare pairs at several separations for each of the two benchmark potentials with thin and thick bubble walls, respectively. We list all parameters in \Cref{tab:n2_runs}. Note that the final time $T$ and simulation volume size $L^3$ are chosen such that each bubble never crosses the box boundaries. This implies the phase transition never actually ends, and the source is still active by the time we compare the spectra.

\begin{table}[t]
  \centering
  \renewcommand{\arraystretch}{1.3} 
  \setlength{\tabcolsep}{4.2pt}
  \begin{tabular}{cccccccc}
  \multirow{2}{*}{$\bar\lambda$} & \multirow{2}{*}{$\gamma_{\rm coll}$} & \multirow{2}{*}{$d$} &
  \multirow{2}{*}{$L$}
    & \multirow{2}{*}{$\Delta x$} & \multirow{2}{*}{$T$}
    & \multicolumn{2}{c}{runtime [GPU-h]} \\
   & & & & & & lattice & (1+1)D \\
  \hline\hline
  $0.84$  & $2.0$ & $27.79$ & $200$ & $0.098$ & $30.18$ & $67.3$ & $0.055$ \\
  $0.84$  & $3.0$ & $41.66$ & $200$ & $0.098$ & $46.27$ & $68.8$ & $0.051$ \\
  $0.84$  & $4.0$ & $55.55$ & $200$ & $0.098$ & $60.35$ & $68.1$ & $0.050$ \\
  \hline
  $0.069$ & $4.0$ & $8.33$  & $50$  & $0.024$ & $9.08$  & $56.7$ & $0.027$ \\
  $0.069$ & $5.0$ & $10.44$ & $50$  & $0.024$ & $11.50$ & $56.8$ & $0.025$ \\
  $0.069$ & $6.0$ & $12.54$ & $50$  & $0.024$ & $13.62$ & $55.9$ & $0.024$ \\
  \end{tabular}
  \caption{Configurations of all two-bubble lattice runs of size $2048^3$ shown in
  \Cref{fig:two-bubble-comparison}.
  $d$ is the bubble separation (used as $R_*$) and $T$ the time at which the spectra are
  compared.
  We also list the total runtime in GPU-hours on Nvidia A100 GPUs and compare it to that of the
  reduced calculation, equally on an Nvidia A100 GPU.}
  \label{tab:n2_runs}
\end{table}

\Cref{fig:two-bubble-comparison} compares the reduced calculation with direct $(3+1)$-dimensional tensor evolution using the same physical energy normalization.
The plotting scale is $R_*=d$, and $\gamma_{\rm coll}$ is the reference contraction label defined in \cref{eq:midpoint-collision-convention}.
The lower panels show the reduced-pair prediction divided by the lattice spectrum after applying the lattice's radial binning to the former.
This uses the same Fourier shells and their discrete mode multiplicities, not an interpolation to the shell centres.
The upper panels retain the original frequency sampling of each calculation.
A ratio is shown only for complete shells within the reduced response's frequency coverage. The error band illustrates the statistical uncertainty on the shell-averaged lattice spectrum, $\sigma_{\mathcal P_{\rm GW}}/\mathcal P_{\rm GW}=\sqrt{2/N_k}$, where $N_k$ is the number of lattice modes in each shell. It does not contain a measure of the numerical uncertainty, which is negligible at low frequencies but grows towards the far tail.

The reduced calculation reproduces the position and amplitude of the spectral peak for both potentials, with agreement extending over much of the falling spectrum.
Larger differences remain in the weak ultraviolet tail of the highest-boost pairs in the displayed data.

\begin{figure*}[t]
    \centering
    \includegraphics[width=\textwidth]{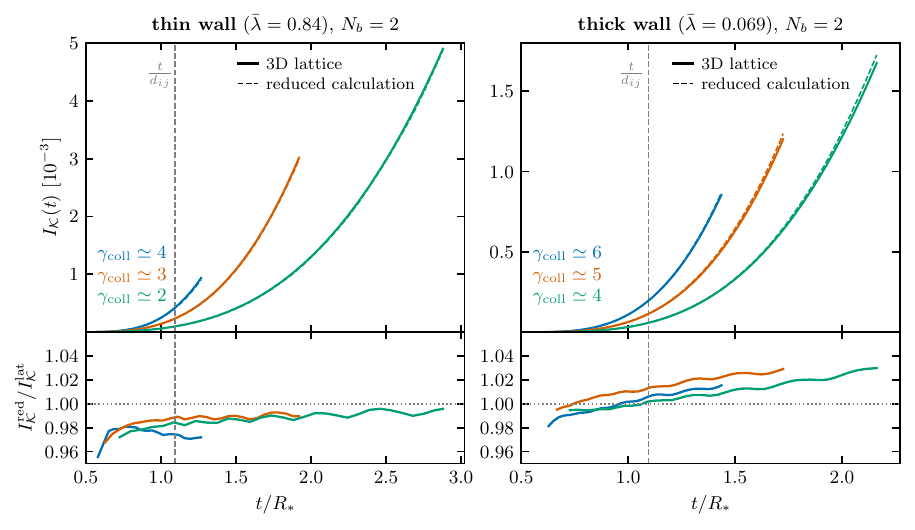}
    \caption{Frequency-integrated gravitational-wave energy for the isolated two-bubble benchmarks of \cref{fig:two-bubble-comparison}.
    The left and right columns show $\bar\lambda=0.84$ and $0.069$; colours identify the same collision boosts as in the spectral comparison.
    Solid curves show the full $(3+1)$-dimensional lattice result and dashed curves show the symmetry-reduced result with the Gaussian cutoff of \cref{eq:pair-window}.
    The upper panels plot $I_{\mathcal K}$ in \cref{eq:integrated-gw-power}, evaluated with the same fixed nonzero Fourier shells for both methods and at all times for each pair, up to the nominal cutoff $kR_*=100$.
    Reduced spectra are evaluated at the occupied lattice-mode radii before shell summation and integration.
    The lower panels show $I_{\mathcal K}^{\rm red}/I_{\mathcal K}^{\rm lat}$ where both energies exceed $10^{-3}$ of that pair's maximum displayed lattice value, excluding ratios of negligible signals.
    The vertical dashed lines mark the reference time $t/d=1.095$ shown in \cref{fig:two-bubble-comparison}.
    No integrated uncertainty band is inferred from the spectral mode-counting estimates, since the correlations between different shells are not measured.}
    \label{fig:two-bubble-integrated}
\end{figure*}

A single spectrum is not sufficient to test the late-time response.
For a fixed frequency interval $\mathcal K=[k_{\min},k_{\max}]$, we use
\begin{equation}
    I_{\mathcal K}(t)=
    \int_{k_{\min}}^{k_{\max}}\mathrm d\ln k\,
       \mathcal P_{\rm GW}(k,t).
    \label{eq:integrated-gw-power}
\end{equation}
\Cref{fig:two-bubble-integrated} follows this energy for the same six pairs, with time expressed in units of $R_*=d$.
Both calculations use the same fixed set of nonzero Fourier shells throughout each history, with a nominal upper cutoff $kR_*=100$.

The direct and reduced histories track one another through the growth of the tensor energy.
At $t/d=1.095$, the reduced-to-lattice ratios span approximately $0.974$--$1.013$ across the six pairs.
At their last displayed snapshots, which occur at different $t/d$, the ratios span approximately $0.972$--$1.030$.
Together, the spectral and integrated comparisons show that the prescription reproduces both the main frequency dependence and the time evolution of the gravitational-wave energy in the selected frequency range.
The quoted differences describe these benchmarks, rather than a numerical convergence error. 
Most importantly, the reduced calculation is about 1200 to 2300 times faster than the full lattice simulation, a cost saving that translates directly to the pairwise surrogate.

\section{Pairwise surrogate: construction and validation}
\label{sec:surrogate-model}
\label{sec:reference-comparison}

The two-bubble calculations resolve what happens when two vacuum walls collide, but a complete transition contains many overlapping collisions.
Our approximation is to use those isolated collisions as building blocks, with a separate geometric calculation determining how much of each pair remains unaffected by the other bubbles.
This separates two questions: how the scalar field evolves in a collision, and how that collision is interrupted by its surroundings.
The first is supplied by the response library tested in \cref{sec:pair-comparison}; the second is modelled by the weights introduced below.

\subsection{From an isolated pair to a bubble configuration}
\label{sec:surrogate-construction}

Consider $N_b$ identical bubbles nucleated simultaneously. For each pair $i,j$, the separation $d_{ij}$ determines the collision boost and selects the corresponding two-bubble response from the library. 
We build the many-bubble spectrum from these pair responses, using time-dependent geometric weights to account for how the surrounding bubbles cover each collision region.
The geometry is particularly simple for simultaneous nucleation: two expanding spherical wall surfaces intersect on a circle.
A third bubble can cover an arc of this circle, leaving the rest exposed.
We define the geometric weight $\mathcal W_{ij}(t)$ as the fraction of its circumference outside all other bubble interiors.
It is zero before contact, unity for a completely exposed circle, and decreases as more of the circle is covered by other bubbles.
The circles and covering surfaces follow the measured position of the energetic wall, located by the maximum of its spatial-gradient energy density.
This is important for thick profiles, whose energetic wall need not follow a fixed field contour.
The radius measurement and geometric construction are detailed in \cref{app:weights}.

The weight is applied to changes in the pair spectrum as the observation time is advanced, rather than multiplying its final spectrum by one surviving fraction: 
\begin{equation}
    \mathcal{P}_\text{GW}^\text{sur}(k,T) =\frac{3}{8\pi G\rho_{\rm vac}^2R_*^2V} \sum_{i<j}\int_0^{T} \mathrm{d}t\,
       \mathcal W_{ij}(t)\,
       \frac{\mathrm d\mathcal E_{2,ij}^\text{red}(k,t)}{\mathrm{d} t}\,.
    \label{eq:surrogate-physics}
\end{equation}
Here $\mathcal E_{2,ij}^\text{red}(k,t)$ is the isolated-pair response from \cref{eq:pair-energy-definition}. We compute the instantaneous response $\mathrm{d}\mathcal E_{2,ij}^\text{red}(k,t)/\mathrm{d}t$ numerically by extracting $\mathcal E_{2,ij}^\text{red}(k,t)$, using the window of \cref{eq:pair-window} with the cutoff half-height time set to different $t$.
The pair contributions are summed incoherently, so interference between different pairs is omitted.
The effective weight $\mathcal W_{ij}$ specifies how a covered region ceases to contribute.
We will consider two possibilities, instantaneous screening $\mathcal W^{\rm inst}_{ij}$ and a gradual screening $\mathcal W^{\rm eff}_{ij}$ introduced below.
In practice, \cref{eq:surrogate-physics} weights the change in each pair spectrum between successive observation times.
A change can be negative when radiation from different stages of the same pair collision interferes destructively.
These are changes in the total spectrum, not independent amounts of radiation emitted during each interval.

The surrogate requires $\mathcal E^\text{red}_{2, ij}(k, t)$ for every pair $ij$ and every observation time $t$ that occurs in a particular realization. 
We tabulate the isolated-pair response on a grid in $\gamma_{\rm coll}$ and $t$, and interpolate in between them. 
For a library covering configurations up to $\gamma_*\le\gamma_{*,\max}$, the boosts are sampled from $\gamma_{\rm coll}=1$ to $2.5\,\gamma_{*,\max}$ in steps of $\Delta\gamma_{\rm coll}=0.4$. 
In principle, even larger $\gamma_{\rm coll}$ collisions are possible, but the 2.5 factor is chosen such that these pairs carry less than $0.3\%$ of the gravitational-wave power. 
Pairs with larger boosts that are outside of our scanning range are assigned the response of the largest tabulated boost. 
Similarly, each pair is evolved from its collision up to $T=2.5\,R_*(\gamma_{*,\max})$, long enough to contain the gradually screened contributions, with the spectrum recorded at intervals of $\Delta t\simeq0.29\,r_\text{mid}$, where $r_\text{mid}$ is the radius of the critical bubble applicable also in the thick-wall case. 
We have verified that both $\Delta\gamma_{\rm coll}$ and $\Delta t$ are fine enough to provide a stable prediction. The 32 frequencies are spaced logarithmically such that $1\le kR_*\le100$ is covered for every $\gamma_*\le\gamma_{*,\max}$. 
Between grid points we interpolate linearly in $\ln\mathcal E_2^\text{red}$, both in $\gamma_{\rm coll}$ and in $t$. 
Once built, such a library reconstructs the spectrum of any configuration with $\gamma_*\le\gamma_{*,\max}$ at negligible computational cost.

\subsection{Gradual relaxation after screening}
\label{sec:weight-decay}

An additional bubble wall does not instantly erase the scalar gradients already present in a pair's collision region.
We therefore allow a newly covered fraction to continue contributing while its weight decays, taking instantaneous screening as a limiting prescription.

To specify this decay, we make three assumptions.
First, its dependence on the scalar potential and pair boost is determined by measuring the gradient relaxation in an isolated two-bubble collision, encoded in the function $\widehat D(t/d_{ij};\bar\lambda,\gamma_{ij})$.
Second, the additional suppression associated with the third wall is the same function of the retained gradient fraction $\widehat D$ for every potential and pair boost.
We implement this assumption by taking the decay kernel to be $\widehat D^{\,c}$, with the same exponent $c=3.6$ for all multi-bubble configurations.
Third, each newly covered fraction starts this decay at its screening time: we neglect further dependence on the age of the pair at that encounter and on the incidence angle and boost of the additional wall. 
The isolated-pair measurements thus supply the microphysics-dependent relaxation history, while a common exponent $c$ describes the additional suppression after screening.
Our many-bubble simulations support the use of this universal relaxation prescription across the potentials and boosts studied here.
The calibration and comparisons are presented in \cref{app:decay-calibration}.

When part of a pair's collision circle is covered by another bubble, we let its contribution fade rather than disappear immediately.
The factor $K_{ij}(t-u)$ specifies how much of its geometric weight remains at time $t$ after screening at time $u$.
We take
\begin{equation}
    K_{ij}(\tau)
    =\widehat D\!\left(\frac{\tau}{d_{ij}};
                        \bar\lambda,\gamma_{ij}\right)^c,
    \qquad \tau=t-u,
    \label{eq:weight-retention-kernel}
\end{equation}
where the time since screening is measured in units of the pair separation $d_{ij}$.
Since $K_{ij}(0)=1$, a newly covered fraction initially keeps its full weight, which then decreases according to the pair's gradient-relaxation history.
The exposed fraction continues to contribute fully.
Adding it to the decaying covered contributions gives the effective weight $\mathcal W^{\rm eff}_{ij}$ used in \cref{eq:surrogate-physics}.
The construction of this weight is detailed in \cref{app:weights}.
If not mentioned otherwise, all surrogate reconstructions will use $\mathcal W^{\rm eff}_{ij}$.

\subsection{Three bubbles: the first reconstruction test}
\label{sec:three-bubble-closure}

\begin{figure*}[t]
    \centering
    \includegraphics[width=\textwidth]{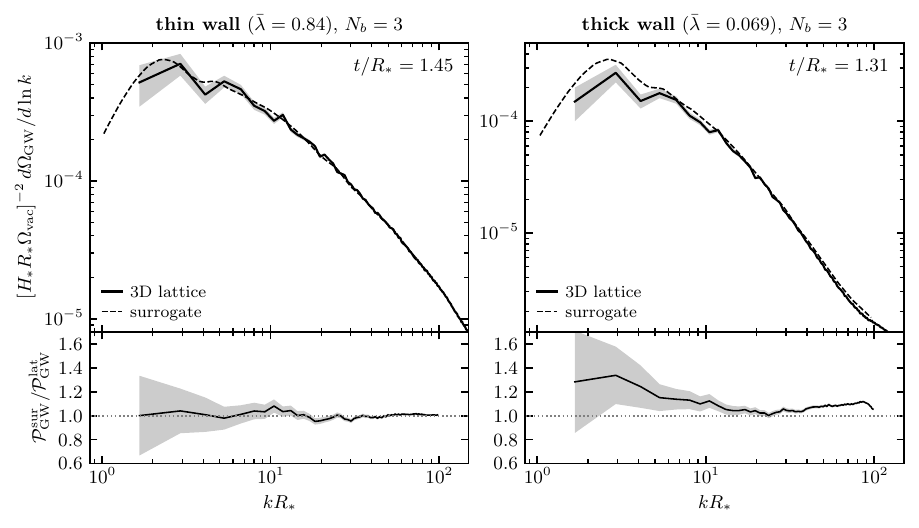}
    \caption{Gravitational-wave spectra for the three-bubble benchmarks $\bar\lambda=0.84$ (left) and $0.069$ (right).
    Solid black curves show the full $(3+1)$-dimensional lattice spectra.
    Dashed curves show the pairwise reconstruction of \cref{eq:surrogate-physics}.
    The lattice snapshots are at $Mt\simeq60.352$ and $13.623$, respectively.
    Here $R_*=(d_{01}+d_{02}+d_{12})/3$.
    Lower panels show $\mathcal P_{\rm GW}^{\rm sur}/\mathcal P_{\rm GW}^{\rm lat}$ with the lattice's radial shell-sum estimator applied to the surrogate, while the upper panels retain their original sampling.
    The normalization is given in \cref{eq:main-spectrum-normalization}.
    Shading shows the same approximate mode-counting estimate as in \cref{fig:two-bubble-comparison}. }
    \label{fig:three-bubble-comparison}
\end{figure*}

Three bubbles provide the simplest test that goes beyond the isolated pair.
Each of the three pair responses is known, but the third bubble interrupts the other two collisions and breaks their individual boost symmetries.
The surrogate must describe this common geometry using the independently computed pair spectra and their screening weights.

For this setup, we reuse the three two-bubble configurations of \Cref{sec:pair-comparison}, now with all three bubbles present in the same box. 
This requires only one additional lattice simulation per $\bar\lambda$, while the surrogate is built directly from the already existing reduced calculations of the three pairs. 
We compare the spectra at time $T$ of the pair with the largest $\gamma$, which by construction is early enough that no collisions across the periodic boundaries have occurred. 
All parameters are listed in \Cref{tab:n3_runs}. 

\begin{table}[t]
    \centering
    \renewcommand{\arraystretch}{1.3}
    \setlength{\tabcolsep}{5pt}
    \begin{tabular}{ccccccc}
    \multirow{2}{*}{$\bar\lambda$} & \multirow{2}{*}{$R_*$} & \multirow{2}{*}{$L$} &
    \multirow{2}{*}{$\Delta x$} & \multirow{2}{*}{$T$}
    & \multicolumn{2}{c}{runtime [GPU-h]} \\
    & & & & & lattice & (1+1)D \\
    \hline\hline
    $0.84$  & $41.67$ & $200$ & $0.098$ & $60.35$ & $67.2$ & $0.156$ \\
    \hline
    $0.069$ & $10.43$ & $50$  & $0.024$ & $13.62$ & $50.4$ & $0.076$ \\
    \end{tabular}
    \caption{Configurations of the three-bubble lattice runs of size $2048^3$ shown in
    \Cref{fig:three-bubble-comparison}.
    The three bubble pairs are those of the two-bubble runs in \Cref{tab:n2_runs}; $R_*$ is their
    mean separation
    and $T$ the time at which the spectra are compared. The runtime of the reduced calculation is
    the sum over the
    three pairs, each on a single Nvidia A100 GPU; the lattice runtime is in GPU-hours on Nvidia
    A100 GPUs.}
\label{tab:n3_runs}
\end{table}

\Cref{fig:three-bubble-comparison} compares this reconstruction with the full $(3+1)$-dimensional lattice result.
For $\bar\lambda=0.84$, it follows the main peak and much of the falling spectrum closely.
The $\bar\lambda=0.069$ benchmark also reproduces the main spectral features, but shows a larger excess around the peak.
These comparisons test the combined radiation and geometric prescriptions at the displayed times.
Together, these benchmarks show that isolated-pair responses combined with geometric screening capture the principal spectral features of three-bubble collisions.
The three reduced pair calculations lower the GPU cost by factors of approximately 430 and 660 relative to the full lattice evolution for $\bar\lambda=0.84$ and $0.069$, respectively.

\subsection{Many-bubble comparisons}
  \begin{table}[!ht]
  \centering
  \renewcommand{\arraystretch}{1.3}   
  \setlength{\tabcolsep}{4pt}         
  \begin{tabular}{ccccccccc}
  \multirow{2}{*}{$\bar\lambda$} & \multirow{2}{*}{$\gamma_*$} & \multirow{2}{*}{$N_b$} &
  \multirow{2}{*}{$L$}
    & \multirow{2}{*}{$R_*$} & \multirow{2}{*}{$\Delta x$}
    & \multirow{2}{*}{$\dfrac{t_{\rm end}}{R_*}$} & \multirow{2}{*}{$\dfrac{T}{R_*}$} & runtime
  \\
   & & & & & & & & [GPU-h] \\
  \hline\hline
  $0.845$ & $4.0$ & $29$ & $175.53$ & $57.13$ & $0.086$ & $1.26$ & $2.33$ & $139$ \\
  $0.845$ & $5.0$ & $15$ & $176.15$ & $71.43$ & $0.086$ & $1.10$ & $2.16$ & $213$ \\
  $0.845$ & $6.0$ & $15$ & $211.40$ & $85.72$ & $0.103$ & $1.22$ & $2.28$ & $185$ \\
  \hline
  $0.069$ & $4.0$ & $64$ & $33.30$  & $8.32$  & $0.016$ & $1.34$ & $2.40$ & $324$ \\
  $0.069$ & $5.0$ & $34$ & $33.80$  & $10.43$ & $0.017$ & $1.27$ & $2.34$ & $250$ \\
  $0.069$ & $6.0$ & $48$ & $45.56$  & $12.54$ & $0.022$ & $1.28$ & $2.50$ & $311$ \\
  \end{tabular}
  \caption{Configurations of the many-bubble lattice runs of size $2048^3$ shown in
  \Cref{fig:n-bubble-comparison}. $R_* = (L^3/N_b)^{1/3}$ is the characteristic bubble separation and $t_{\rm end}$ the time at which the simulation volume $L^3$ is completely filled by the true vacuum, i.e.\ the end of the phase transition. $T/R_*$ denotes the surrogate evaluation time used in the decay calibration of \cref{app:decay-calibration}. For comparison, building the surrogate takes 1.7 GPU hours for $\bar\lambda=0.069$ and 24.1 GPU hours for $\bar\lambda=0.845$; once built, it reconstructs the spectrum of any realization with $\gamma_*\leq6$ at negligible additional cost.}
  \label{tab:many-bubble-benchmarks}
  \end{table}

\begin{figure*}[t]
    \centering
    \includegraphics[width=\textwidth]{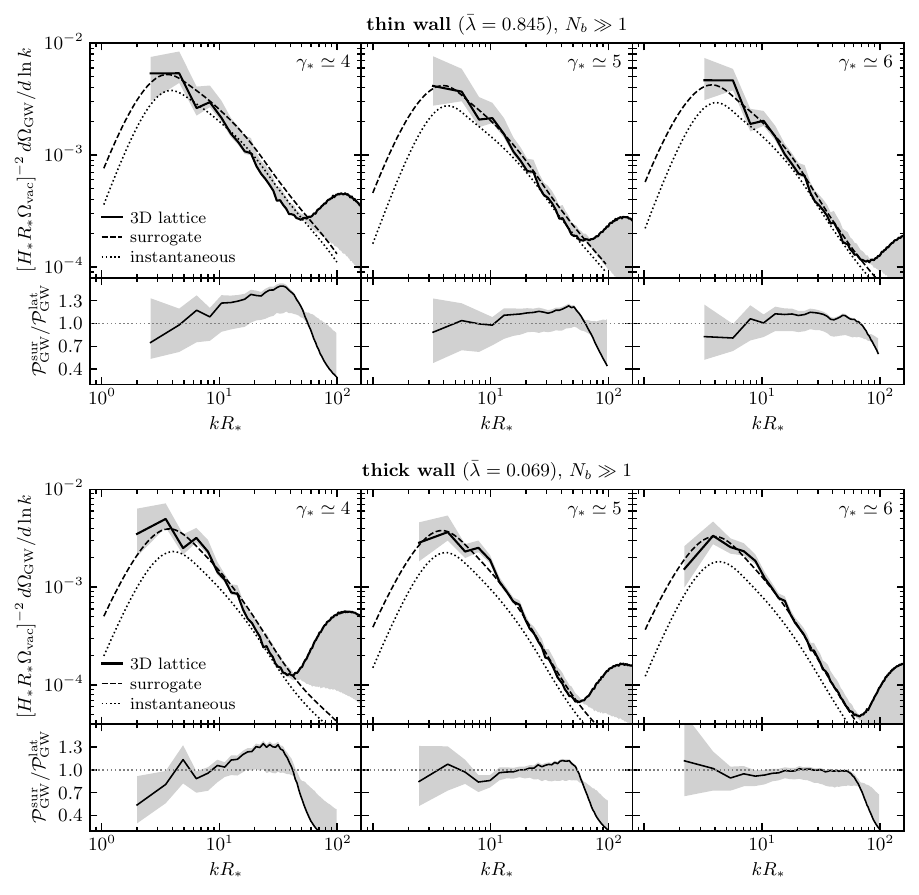}
    \caption{Many-bubble gravitational-wave spectra at characteristic boosts $\gamma_*\simeq4,5,6$ from left to right.
    The upper row shows $\bar\lambda=0.845$ with $N_b=29,15,15$; the lower row shows $\bar\lambda=0.069$ with $N_b=64,34,48$.
    Solid curves show the late-time $(3+1)$-dimensional lattice spectra.
    Gray bands combine variation over the selected late snapshots with an approximate mode-counting estimate.
    Dashed curves use the surrogate model with full post-screening decay; dotted curves, labelled ``instantaneous'', use the surrogate but with instantaneous geometric screening.
    Lower panels divide the dashed surrogate by the lattice result after applying the same radial shell-sum estimator to both; horizontal dotted lines mark unity.
    Upper panels retain the original spectral sampling.
    Here $R_*=L/N_b^{1/3}$ and the normalization is given in \cref{eq:main-spectrum-normalization}.
    Simulation parameters and comparison conventions are given in \cref{tab:many-bubble-benchmarks} and the text.}  
    \label{fig:n-bubble-comparison}
\end{figure*}

\label{sec:many-bubble-validation}
So far we have considered collisions of a few bubbles in an effectively open volume: the walls keep expanding, and the phase transition never completes. 
We now fill a periodic volume with enough simultaneously nucleated bubbles such that the phase transition ends at $t_{\rm end}\simeq1.1$--$1.3\,R_*$ (\cref{tab:many-bubble-benchmarks}). 
We continue the lattice simulations well beyond this point to let the spectrum settle, typically to at least $t/R_*=4$. 
\Cref{fig:n-bubble-comparison} compares the resulting lattice spectra at characteristic boosts $\gamma_*\simeq4$--6 with the surrogate, with and without the gradual relaxation of the collision weights. 
The bubble number and positions differ between the columns, so these are separate benchmarks rather than a controlled sequence in $\gamma_*$. 
For each benchmark, we use the last available lattice spectrum, which is at $t/R_*=8$ for the thick-wall $\gamma_*=4$ and $6$ scenarios and at $t/R_*=4$ for all others.
The shaded bands show their variation over that time, combined in quadrature with the shell sampling uncertainty used in the few-bubble comparisons. 
Before taking a ratio, we apply the lattice's radial binning to the surrogate, as previously detailed in \cref{sec:pair-comparison}.

\begin{figure*}[t]
    \centering
    \includegraphics[width=\textwidth]{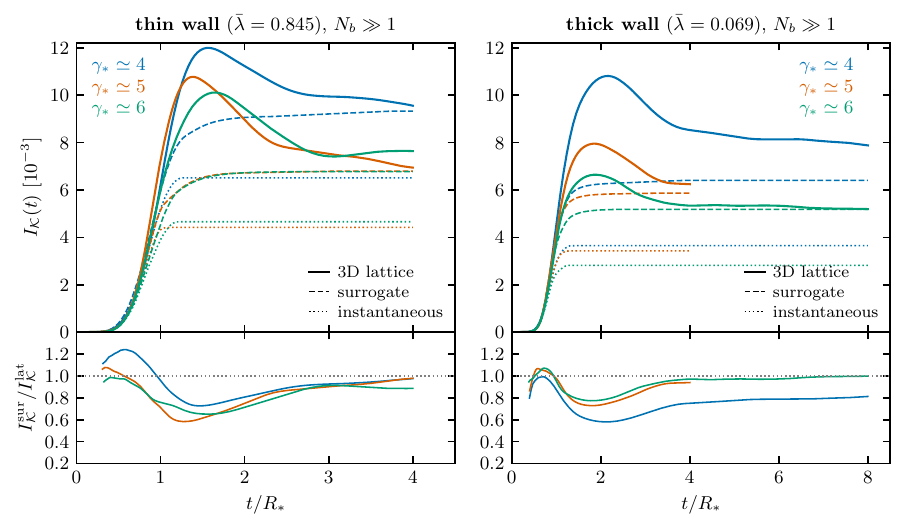}
    \caption{Frequency-integrated gravitational-wave energy for the benchmarks of \cref{fig:n-bubble-comparison}.
    The left and right columns show $\bar\lambda=0.845$ and $\bar\lambda=0.069$; blue, orange and green identify $\gamma_*\simeq4,5,6$.
    The bubble numbers are $N_b=29,15,15$ on the left and $64,34,48$ on the right.
    Upper panels show $I_{\mathcal K}(t)$ from \cref{eq:integrated-gw-power}, summed over fixed nonzero Fourier shells up to the nominal cutoff $kR_*=30$.
    Solid curves show the unwindowed full-energy lattice histories, dashed curves the gradually decaying-weight surrogate, and dotted curves instantaneous screening.
    Lower panels show $I_{\mathcal K}^{\rm sur}/I_{\mathcal K}^{\rm lat}$ for gradual screening.
    Time is normalized by $R_*=L/N_b^{1/3}$.
    The same shell-sum integration estimator is used for both methods, and no uncertainty band is assigned to the integrated ratios.}
    \label{fig:n-bubble-integrated-power}
\end{figure*}
Gradual screening raises the reconstructed signal and improves agreement around the main peak in these examples.
It also follows the broad falling spectrum, although appreciable residuals remain and the improvement is not uniform across all frequencies.
The comparison supports retaining a post-screening contribution, but does not uniquely determine its decay law.

An upturn of the lattice spectra at high frequencies emerges after the phase transition has finished, where the field is left excited and continues to oscillate around the true vacuum. These oscillations continue to source gravitational waves at $kR_*\sim m_b R_*$; thus, the peak location moves further to the UV at larger $\gamma_*$. This oscillation phase is not captured by the reduced calculation, which is expected and not an issue. In a realistic phase transition, this secondary peak sits far in the UV with an amplitude that is well out of range of any future gravitational wave detector. That the surrogate does not reproduce this peak can therefore be considered a feature, not a bug.

At frequencies above the main collision peak, the surrogate reproduces the broad falloff of the lattice spectra in \cref{fig:n-bubble-comparison}.
Agreement along this UV tail improves at larger $\gamma_*$, whereas the surrogate predicts a shallower falloff than the lattice result in both simulations with $\gamma_*\simeq 4$.
The larger discrepancy at smaller boosts may reflect finite-width effects in interactions with additional bubble walls.
Although the isolated-pair simulations resolve the wall profiles, both weight prescriptions describe these additional interactions through the geometric coverage of a single wall surface.
At smaller boosts, the walls are less Lorentz contracted, and their interaction with the existing gradients extends over a wider region, whose dynamics may require a more detailed treatment of screening.

At each time, we integrate the spectra over the same fixed frequency range to obtain $I_{\mathcal K}(t)$, defined in \cref{eq:integrated-gw-power}.
\Cref{fig:n-bubble-integrated-power} compares the resulting gravitational-wave energy as a function of time in the pairwise surrogate and the full $(3+1)$-dimensional lattice simulations.
We show both instantaneous and gradual screening to examine how the treatment of covered collision regions affects the build-up and subsequent evolution of the reconstructed signal.

Both surrogate curves use the same isolated-pair responses and geometric screening histories.
Instantaneous screening removes a covered contribution immediately, while gradual screening allows it to continue contributing with a weight that decays as $\widehat D^{\,c}$.
Only the gradual prescription introduces the common exponent $c$, which we calibrate against the six many-bubble benchmarks in \cref{tab:many-bubble-benchmarks}, adopting $c=3.6$ for both potentials and all configurations.
The calibration is described in \cref{app:decay-calibration}.
The separation between the surrogate curves therefore shows the effect of this common post-screening relaxation prescription.

One of the key differences between the full $(3+1)$-dimensional lattice simulation and the surrogate is the behaviour of $I_\mathcal{K}(t)$ at intermediate times ($1 \lesssim t/R_* \lesssim 2.5$ for thin wall and $1.5 \lesssim t/R_* \lesssim 3$ for thick wall). In the full lattice simulation, the total GW energy grows to a maximum before destructive interference reduces this peak to settle to a late-time value, which is typically in agreement with the surrogate model. This interference occurs only after all the bubble walls of the simulation have collided with one or more neighbouring bubbles and therefore appears to be beyond the dynamics captured by the surrogate. Nevertheless, the aforementioned approach with gradual decay of the weights reproduces the late-time total power in GWs. 

Regarding the computational cost, building the surrogate is roughly 6-9 times faster for $\bar\lambda=0.845$ than a single lattice simulation and 150-190 times faster at $\bar\lambda=0.069$.
Note, however, that this is skewed heavily to the lower end, as we simulate only a relatively small number of bubbles ($N_b\leq 64$) in our lattice simulations. 
These simulations with fewer bubbles are still statistically meaningful and a non-trivial test for our surrogate model, as even $N_b=15$ corresponds to $\sim 115$ individual colliding pairs and $\sim490$ pairs at $N_b=64$. Yet, the number is small compared to existing simulations such as those in~\cite{Cutting:2020nla}, which simulate up to 4096 bubbles. Since the build time of our surrogate does not depend on the number of bubbles in any given realization, and evaluating the surrogate is cheap, the cost savings for these large-$N_b$ simulations are drastically larger.

In addition to our own lattice simulations, we also compare our surrogate model to simulations presented in the literature in \Cref{app:literature}.

\section{Accessing the high-boost regime}
\label{sec:ultrarelativistic-limit}

\begin{figure*}[t]
    \centering
    \includegraphics[width=\textwidth]{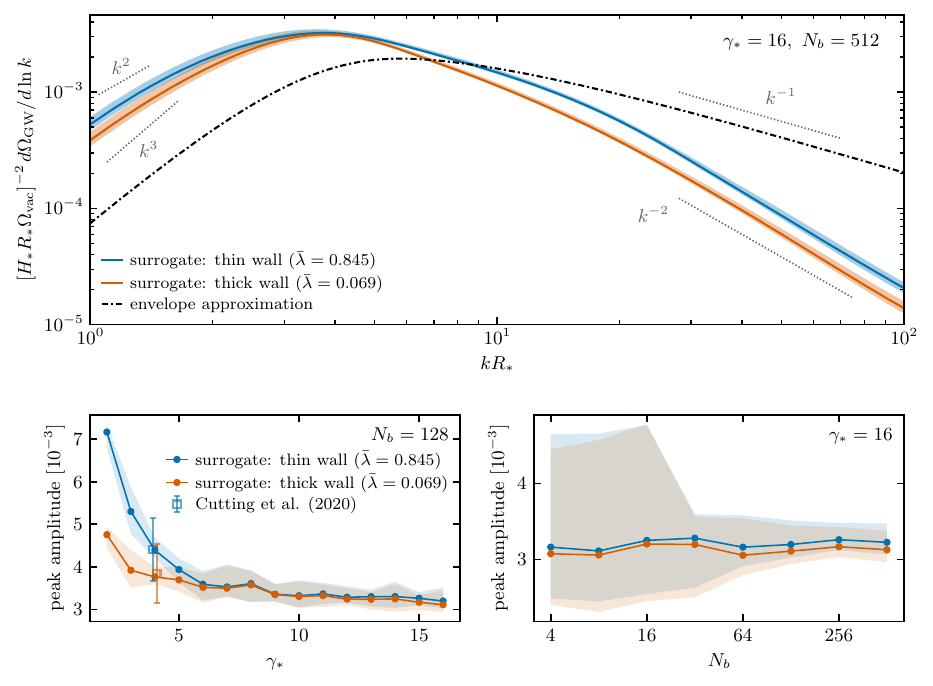}
    \caption{Surrogate gravitational wave spectra and peak amplitudes.
    Blue and orange show $\bar\lambda=0.845$ and $0.069$.
    The upper panel shows the pointwise median spectrum of 16 independent bubble catalogues at $\gamma_*=16$ and $N_b=512$.
    The lower panels show the peak spectral amplitude over the computed frequency interval: $N_b=128$ is fixed while $\gamma_*$ varies on the left, and $\gamma_*=16$ is fixed while $N_b$ varies on the right.
    Central curves give the maximum of the pointwise median spectrum; shading spans the minimum and maximum of the individual realization peaks.
    The bands describe realization spread, not numerical convergence or an uncertainty on the median.
    Here $R_*=L/N_b^{1/3}$ and the normalization is given in \cref{eq:main-spectrum-normalization}.
    The decay prescription for the weights and its calibration entering the surrogate are described in \cref{app:weights}.
    We overlay our results in the lower left panel with those from~\cite{Cutting:2020nla}, which ran full $(3+1)$-dimensional lattice simulations at $\gamma_*=4$ with $N_b=512$ at both $\bar\lambda=0.845$ and $\bar\lambda=0.069$, rendering them directly comparable to our surrogate. We present a more detailed comparison in \Cref{app:literature}.}
    \label{fig:predictions}
\end{figure*}

The pairwise surrogate extends our study of gravitational-wave production to collision boosts beyond the practical reach of direct many-bubble simulations.
We use it to explore how the collision-scale spectrum depends on boost and scalar potential, and how its variation between realizations changes as the number of bubbles increases.
Throughout, we use the same gradual-screening prescription as in the moderate-boost comparisons, with the common relaxation exponent calibrated in \cref{app:decay-calibration}.

For demonstration, we built two surrogate models, for $\bar\lambda=0.845$ and $\bar\lambda=0.069$, that can generate gravitational wave spectra up to $\gamma_*=16$. This requires generating spectra for individual bubble collisions at a boost of up to $\gamma_{ij}=40$, which is well out of reach of any lattice simulation in the immediate future. 
Yet, in our reduced simulation it is possible at an affordable cost. The total runtime was 165 GPU-hours for $\bar\lambda=0.069$ and 709 GPU-h for $\bar\lambda=0.845$, which is comparable to a single lattice simulation at a small boost.

\subsection{Dependence on characteristic boost}
\label{sec:large-boost-sequence}

As bubbles expand, their walls accelerate and become increasingly thin compared with the bubble separation \cite{Cutting:2020nla}.
Both initially thin- and thick-walled bubbles therefore approach a common ultrarelativistic thin-shell geometry before collision.
The underlying scalar potential governs the subsequent field evolution and gradient relaxation, with isolated-pair studies finding distinct collision dynamics and gravitational-wave spectra even at highly relativistic boosts \cite{Gould:2021dpm}.
Here we investigate how this potential dependence persists in many-bubble transitions as the characteristic collision boost increases.

We reconstruct the gravitational-wave spectrum at $\gamma_*=16$ from 16 realizations with $N_b=512$, shown in the upper panel of \Cref{fig:predictions}.
The two potentials, $\bar\lambda=0.845$ and $0.069$, give nearly equal peak amplitudes but different post-peak shapes.
On the low-frequency side, fits over $1\leq kR_*\leq2$ give $\mathcal P_{\rm GW}\propto k^{1.95}$ and $k^{2.14}$ for the thin- and thick-wall cases, respectively.
At high frequencies, $30\leq kR_*\leq100$, both spectra fall approximately as $k^{-2.1}$.
Between the peak and this high-frequency tail, fits over $6\leq kR_*\leq16$ give $\mathcal P_{\rm GW}\propto k^{-1.12}$ for thin walls and $k^{-1.41}$ for thick walls.
The post-peak shape therefore retains sensitivity to the scalar potential even when the peak amplitudes have become nearly equal.

We also compare our results with a semi-analytic envelope spectrum for simultaneous nucleation, following the stress-correlation approach of Ref.~\cite{Megevand:2021features}.\footnote{The envelope spectrum can alternatively be computed numerically using the exact bubble centres of each realization, by integrating over the uncollided wall surfaces; see for example Ref.~\cite{Weir:2016tov}. We have independently tested that both methods agree.}
The envelope prediction has a lower peak amplitude at a higher frequency and a shallower ultraviolet tail, falling approximately as $k^{-1}$. 
This further highlights that ultra-relativistic thin walls alone do not reproduce the envelope result.
The pair library captures the potential-dependent scalar-field evolution after collision, allowing the surviving gradients to continue sourcing gravitational waves \cite{Gould:2021dpm}.

The lower-left panel of \cref{fig:predictions} scans $\gamma_*=2$--16 at fixed $N_b=128$, using 16 independent catalogues at each point. For both $\bar\lambda$ values, the peak amplitude decreases as the boost increases. 
Over $10\leq\gamma_*\leq16$, linear fits weighted by the realization-band widths give decreases in the central peak amplitude of approximately $3.6\%$ for thin walls and $5.8\%$ for thick walls. The peak amplitudes therefore vary only weakly with boost in this range.

\subsection{Dependence on bubble number}
\label{sec:many-bubble-application}

The bubble-number scan in the lower-right panel of \cref{fig:predictions} keeps $\gamma_*=16$ fixed and varies $N_b$ from 4 to 512 using 16 realizations each. 
At small $N_b$, an individual collision geometry can dominate the spectrum; a larger ensemble samples more of the screening geometries that occur in a transition.
As a result, the median peak amplitude is relatively stable, but the scatter, illustrated by the envelope band in \cref{fig:predictions}, decreases as the number of bubbles increases. 
At $N_b=512$, the realization band extends from approximately $5\%$ below to $8\%$ above the central peak amplitude for both potentials.

\subsection{The origin of the gravitational wave power}
\label{sec:origin-of-power}

\begin{figure*}[t]
    \centering
    \includegraphics[width=\textwidth]{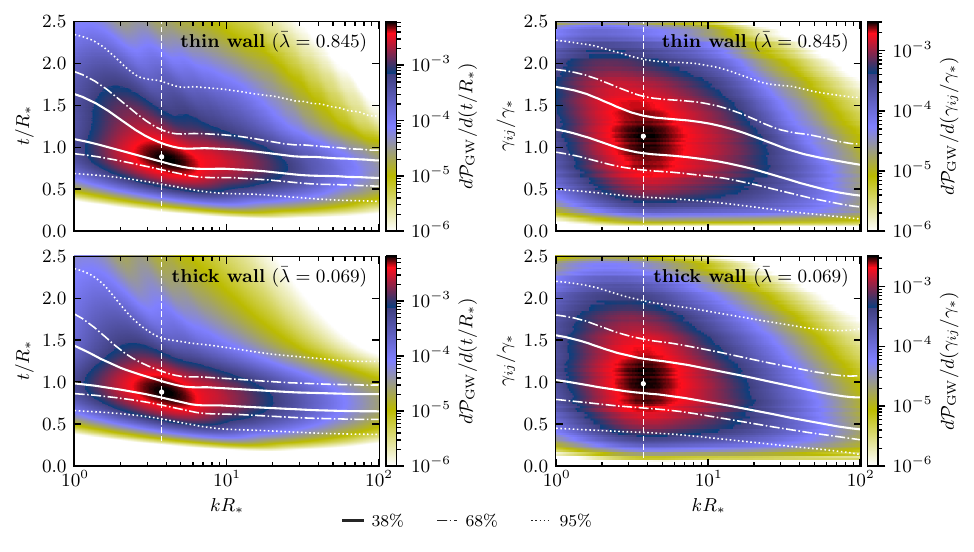}
    \caption{Origin of the gravitational-wave power in the surrogate spectrum for $\gamma_*=16$ and $N_b=512$, averaged over 16 realizations of the bubble positions. The colour shows how the spectrum $\mathcal{P}_{\rm GW}(k)$ is built up as a function of the time $t/R_*$ at which the power is sourced (left) and of the Lorentz factor $\gamma_{ij}/\gamma_*$ of the colliding pair that sources it (right), for thin (top, $\bar\lambda=0.845$) and thick walls (bottom, $\bar\lambda=0.069$). At each $kR_*$ the lines enclose the central 38\% (solid), 68\% (dash-dotted) and 95\% (dotted) of the power. The dashed vertical line marks the peak of the spectrum, $kR_*\simeq3.7$, and the white dot marks the maximum of the distribution at that wavenumber. }
    \label{fig:power-origin}
\end{figure*}

Another advantage of the surrogate model is that we can trivially disentangle the origin of the gravitational wave power by keeping track of which bubble pairs contribute to \cref{eq:surrogate-physics} and at which stage. This is not directly possible in a lattice simulation, as all of the radiation is measured simultaneously. We present the differential spectrum as a function of the normalized collision time $t/R_*$ and the normalized bubble pair collision boost $\gamma_{ij}/\gamma_*$, based on our realizations at $\gamma_*=16$. We present normalized quantities, as the result is qualitatively similar at other $\gamma_*$ values, within the range available from the surrogate.

For the collision time, we find that the peak emission is dominated by radiation during a relatively narrow window around $t/R_*=1$. Interestingly, the UV tail develops slightly earlier at $t/R_*$ between 0.5 and 1, whereas the IR tail is the opposite, with most contribution coming at $t/R_*$ between 1.0 and 1.5. As for the individual bubble collisions' boosts $\gamma_{ij}$, the peak is made up more broadly, with most contributions from bubble pairs with $\gamma_{ij}/\gamma_*$ between 0.7 and 1.4. The UV tail is dominated by smaller $\gamma_{ij}$, whereas the IR tail is dominated by larger $\gamma_{ij}$. 2
Diagnostics like this can help us understand the dynamics behind a phase transition more closely, as they shed light on the internal mechanisms that are otherwise obscured.  

\section{Discussion and conclusions}

We have developed a pairwise surrogate for gravitational-wave production in vacuum phase transitions.
The nonlinear scalar dynamics are computed in a reusable library of symmetry-reduced $(1+1)$-dimensional two-bubble simulations (see \cref{sec:two-bubble-setup}).
The pair spectra are then combined in an incoherent sum, with time-dependent geometric weights accounting for screening by other bubbles, as described in \cref{sec:surrogate-construction}.
This construction retains the post-collision evolution of the scalar gradients while avoiding the need to resolve every Lorentz-contracted wall throughout a three-dimensional volume.

For isolated two-bubble collisions, the comparisons with full $(3+1)$-dimensional lattice simulations in \cref{fig:two-bubble-comparison,fig:two-bubble-integrated} show that the reduced radiation calculation, including its finite-duration cutoff, reproduces the spectral peaks and the evolution of the frequency-integrated gravitational-wave energy.
Moving to three bubbles provides the first test of the surrogate model itself.
The resulting reconstruction captures the principal spectral features of the full three-bubble lattice benchmarks in \cref{fig:three-bubble-comparison} for both thin and thick bubble walls.

The many-bubble comparisons highlight the importance of scalar gradients that continue sourcing gravitational waves after their collision region is covered by another bubble.
The gradual-screening prescription of \cref{sec:weight-decay} includes this contribution and brings the surrogate into closer agreement with the lattice spectral peaks and frequency-integrated energy in \cref{fig:n-bubble-comparison,fig:n-bubble-integrated-power}.
The same prescription gives this improvement across both scalar potentials and the characteristic boosts studied.

The surrogate has two main advantages: First, it is vastly cheaper to build and evaluate. 
In the isolated-pair benchmarks of \cref{tab:n2_runs}, the reduced calculations require approximately $1200$--$2300$ times fewer GPU-hours than their full lattice counterparts.  
Once built, a library covering a given potential and pair-boost range can be reused across realizations of the bubble positions.
This enables ensemble calculations with thousands of bubbles and extends the accessible collision boosts far beyond those of direct many-bubble lattice simulations. The second advantage is control over the full phase transition history. It makes it possible to disentangle different contributions to the gravitational wave spectrum and isolate the dominant ones. It gives us a unique insight into the dynamics of such phase transitions. 

Using this approach, the applications in \cref{sec:ultrarelativistic-limit} reach characteristic boosts of $\gamma_*=16$ and configurations containing up to $N_b=512$ bubbles.
The pair library extends to individual collision boosts of $\gamma_{ij}=40$, compared with $\gamma_{ij}=16$ in the isolated-pair gravitational-wave calculations of Ref.~\cite{Gould:2021dpm}.
Our largest characteristic boost is approximately twice the largest nominal value reached in the direct many-bubble lattice simulations of Ref.~\cite{Cutting:2018tjt}.
At fixed potential and bubble number, doubling the characteristic boost would increase the estimated cost of a direct uniform-lattice simulation by a factor of $2^8=256$, maintaining the same resolution of the contracted walls and evolution interval in units of $R_*$.
This follows from the $\gamma_*^8$ cost scaling discussed in \cref{sec:introduction}.

The boost scan in the lower-left panel of \cref{fig:predictions} shows that the normalized peak amplitude decreases as the characteristic boost increases from $\gamma_*=2$ to $16$.
The decrease is strongest at lower boosts, while the peak amplitude varies by only a few per cent over $10\leq\gamma_*\leq16$ for both potentials.
The difference between the two potentials also narrows as the boost increases.
At $\gamma_*=16$, $\bar\lambda=0.845$ and $0.069$ give nearly equal peak amplitudes while retaining distinct post-peak slopes, as shown in the upper panel of \cref{fig:predictions}.
The spectral shape therefore continues to distinguish the two potentials even when their peak amplitudes have become nearly equal.
Both spectra fall more steeply at high frequencies than the envelope prediction.
Increasing the number of bubbles reduces the realization-to-realization scatter while leaving the central peak amplitude comparatively stable, as illustrated in the lower-right panel of \cref{fig:predictions}.
The time-resolved reconstruction in \cref{fig:power-origin} further identifies the collision stages and pair boosts that build up different parts of the spectrum.
The peak develops mainly around $t/R_*\simeq1$, with higher-frequency components developing earlier and lower-frequency components later.

These results have several implications for the gravitational-wave signal of vacuum transitions. First, a substantial part of the spectrum is sourced by scalar gradients that continue to radiate after their collision region has been covered by another bubble. Neither the envelope approximation nor a pure thin-shell description captures this contribution, which explains why the many-bubble spectra differ from the envelope prediction even when the walls are highly Lorentz-contracted.

Second, the boost scan in the lower-left panel of \cref{fig:predictions} indicates that the spectrum approaches a large-boost regime. The normalized peak amplitude decreases most strongly at low boost and varies by only a few per cent over $10\leq\gamma_*\leq16$ for both potentials, while the difference between the two potentials narrows.
If this trend continues, spectra computed at $\gamma_*\sim\mathcal O(10)$ already capture the peak amplitude of transitions with much larger boosts, although our results do not establish the asymptotic limit.

Third, the scalar potential remains imprinted on the spectral shape even where it no longer affects the peak amplitude. At $\gamma_*=16$, $\bar\lambda=0.845$ and $0.069$ give nearly equal peak amplitudes but distinct post-peak slopes, $k^{-1.1}$ and $k^{-1.4}$, before both approach $k^{-2.1}$ at high frequencies. A single universal spectral template is therefore insufficient for vacuum transitions; the intermediate slope carries information about the scalar potential.

Finally, the time-resolved reconstruction in \cref{fig:power-origin} shows that the peak is built up mainly around $t/R_*\simeq1$ by pairs with $\gamma_{ij}/\gamma_*\simeq0.7$--$1.4$, that is, by typical collisions during the transition rather than by late collective dynamics.
This supports the premise of the pairwise surrogate and identifies the collision stage that a phenomenological model of the peak must describe.
Increasing the number of bubbles reduces the realization-to-realization scatter while leaving the central peak amplitude comparatively stable, as illustrated in the lower-right panel of \cref{fig:predictions}.

An important future direction is to include time-dependent nucleation histories, allowing bubbles of different ages and sizes to collide.
For ideal vacuum bubbles with spacelike-separated nucleation events, a boost along the pair axis makes nucleation simultaneous and preserves the $(1+1)$-dimensional scalar evolution \cite{Kosowsky:1992collisions,Gould:2021dpm}.
The reconstruction would then require geometric weights based on each bubble's age and radiation integrals with the cutoff defined in the common transition frame.
In addition, the direction-dependent radiation must be transformed to that frame before angular averaging.
We leave this extension and its comparison with unequal-time lattice simulations to future work.

\begin{acknowledgments} 
{\it 
We thank Aleksandr Azatov, Salvatore Bottaro, Mark Hindmarsh, Thomas Konstandin, Mehrdad Mirbabayi, Diego Redigolo and Lorenzo Ubaldi for insightful discussions. In particular, we thank Joshua W. Foster and Sarah R. Geller for their contributions to the initial ideas and early development of this project and David Weir for invaluable correspondence.

The physical model, methods and original simulation code were developed by the authors. Large language models were used to assist with parts of the code implementation and optimization, analysis, and plotting scripts, under the direction of the authors. All code and results were checked and validated by the authors, who take full responsibility for the content of this paper.

The authors wish to acknowledge the Institute for Fundamental Physics of the Universe (IFPU), in Trieste, for hosting and supporting this work through their team research program. TO also acknowledges the hospitality and partial support of the Galileo Galilei Institute (GGI) in Florence, where part of this work has been carried out during a team research program.

This research used resources of the National Energy Research Scientific Computing Center (NERSC), a U.S. Department of Energy Office of Science User Facility located at Lawrence Berkeley National Laboratory, operated under Contract No. DE-AC02-05CH11231 using NERSC award HEP-ERCAP 0023978.
}
\end{acknowledgments}

\section*{Data availability}
The code used in this work, the gravitational-wave spectra of the lattice simulations, and the data underlying all figures are available at \url{https://github.com/MSABuschmann/PhaseTransitionBuilder}. The lattice simulations were performed with \textsc{Sledgehamr}; the implementation of the scalar and tensor evolution used here is available in its repository~\cite{sledgehamr_code}. The full pair-spectrum libraries and geometric-weight catalogues used for the large-boost predictions are available from the authors upon reasonable request.

\bibliography{refs}
\clearpage
\appendix
\onecolumngrid
\begin{figure*}
    \centering
    \includegraphics[width=0.95\textwidth]{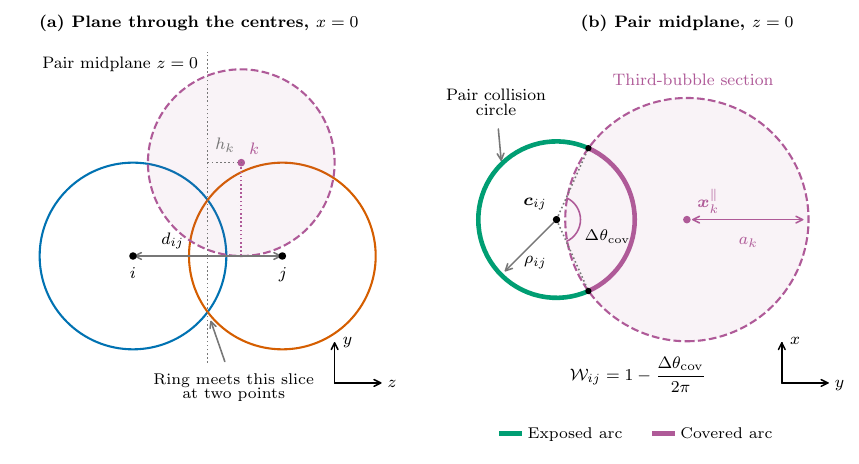}
    \caption{Geometric screening of a pair collision by a third bubble, shown in two orthogonal spatial sections of the same equal-radius configuration.
    Left: the plane through the three bubble centres, with the pair axis along $z$.
    The two spherical walls intersect on a circle in their midplane $z=0$; this section meets that circle at only two points.
    Right: the pair midplane, showing the complete collision circle with centre $\boldsymbol c_{ij}$ and radius $\rho_{ij}$.
    The third bubble cuts a disk of radius $a_k$ in this plane, centred on the projection $\boldsymbol x_k^{\parallel}$ of its centre; $h_k$ is its perpendicular distance from the plane.
    The purple arc lies inside that disk and is screened, while the teal arc remains exposed.
    The weight is the exposed fraction of the circumference, not the uncovered area.}
    \label{fig:weight-geometry}
\end{figure*}

\twocolumngrid
\section{Geometric weights and relaxation prescription}
\label[appendix]{app:weights}

The surrogate needs two inputs for each bubble pair: its isolated collision spectrum and a weight describing how the surrounding bubbles interrupt that collision.
This appendix explains how we construct the second input.
We first define a wall radius and use it to find the exposed fraction of each collision circle.
We then measure gradient relaxation in an isolated collision to set the time dependence of a screened contribution.
Finally, we combine the exposed and decaying fractions into the weight applied to successive changes in the pair spectrum.

\subsection{Wall radius and screening geometry}

The geometric calculation represents each bubble wall by a sphere whose radius follows the energetic part of the scalar profile.
We measure that radius on the undisturbed side of a two-bubble $(1+1)$-dimensional simulation using
\begin{equation}
    R_g(t)=\operatorname*{arg\,max}_{r}
              [\partial_r\phi(t,r)]^2\,.
    \label{eq:gradient-wall-radius}
\end{equation}
It need not coincide with the analytic half-amplitude radius $R_{\mathrm{mid}}(t)$ in \cref{eq:wall-trajectory}, especially for thick profiles.
The associated wall width is the full width at half maximum of the gradient peak, not the separation of fixed field contours in \cref{eq:wall-contours}. This choice follows the part of the wall that sources gravitational waves: the source in \cref{eq:source-fourier-integral} is built from $\partial_i\phi\,\partial_j\phi$, so the wall position relevant for the radiation is where the gradient energy peaks. The fixed field contours are set by the profile at nucleation and, for thick profiles, can drift into regions that carry little gradient energy once the interior field relaxes. The measured radius enters only the weights; the pair spectrum assigned to each collision is still selected by its separation $d_{ij}$.

\begin{figure*}[!t]
    \centering
    \includegraphics[width=\textwidth]{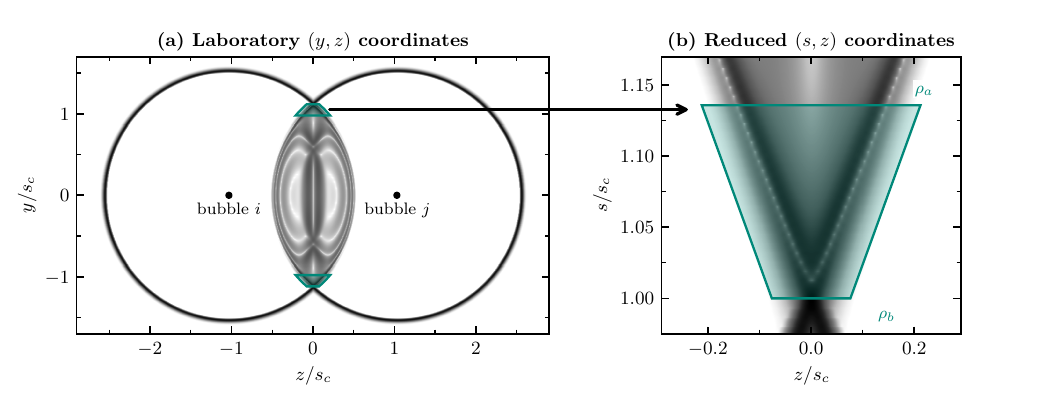}
    \caption{Gradient-retention sampling volume for $\bar\lambda=0.84$ and $\gamma_{\rm coll}=4$, shown at $t=t_{\rm ref}=1.5s_c$ in two coordinate systems.
    Left: a section through the collision axis $z$, with transverse coordinate $y$ and $\rho=|y|$; rotating the upper teal region about $z$ gives $\mathcal P(t_{\rm ref})$.
    Right: the same sampling region expressed in the reduced $(s,z)$ coordinates, using $s=\sqrt{t_{\rm ref}^2-\rho^2}$, with enlarged axes.
    The outer and inner annular radii map to its lower and upper edges, respectively.
    Both panels refer to the same laboratory time: different values of $s$ on the right correspond to different transverse radii, not different observation times.
    Darker shading indicates larger spatial gradient-energy density on this laboratory-time slice, using the same grey scale in both panels; the outer wall caps on the left use the undisturbed bounce profiles. }
    \label{fig:gradient-retention-patch}
\end{figure*}

For simultaneously nucleated identical bubbles, the two spheres have the same  $R_g(t)$.
As shown in \cref{fig:weight-geometry}, their intersection after contact is a circle in the plane halfway between their centres, perpendicular to the pair axis.
Its centre and radius are
\begin{equation}
    \boldsymbol c_{ij}=\frac{\boldsymbol x_i+\boldsymbol x_j}{2},
    \qquad \rho_{ij}(t)=\sqrt{R_g(t)^2-\frac{d_{ij}^2}{4}}.
    \label{eq:collision-circle-geometry}
\end{equation}
An additional bubble whose centre is a distance $h_k$ from the pair midplane cuts a disk of radius $a_k=\sqrt{R_g^2-h_k^2}$ in that plane, provided $|h_k|<R_g$.
The arc of the collision circle inside this disk is screened, as illustrated by the purple segment in \cref{fig:weight-geometry}.
For each colliding pair, we iterate over all other bubbles in the simulation volume, carefully taking into account periodic boundary conditions. We combine all covered arc sections for that pair, counting overlapping arcs only once, and divide the remaining arc length by $2\pi\rho_{ij}$ to obtain $\mathcal W_{ij}$. This way, all weights $\mathcal W_{ij}$ will drop naturally to zero once the phase transition has finished. We evaluate $\mathcal W_{ij}$ on a uniform time grid with spacing $R_*/450$, from nucleation until 5\% after the simulation volume is completely filled.

Related statistical approaches describe screening through uncollided-wall probabilities and their correlations \cite{Megevand:2020correlations}; these are distinct from the circumference weights used here.

\subsection{Gradient-retention measurement}

Scalar gradients left behind by a wall collision can continue to source gravitational waves.
We use their subsequent evolution to inform how a screened contribution should decay.
Rather than integrate over the entire growing collision region, we measure the gradient energy in a narrow annulus at a fixed distance from the collision axis.
This samples a limited portion of the colliding walls, while allowing its extent along the axis to grow to include the outgoing disturbances.

We make this measurement in an isolated pair simulation.
We place the origin halfway between its centres and take the $z$ axis along the line joining them, so that the centres are at $z=\pm d/2$.
In cylindrical coordinates, $\rho=\sqrt{x^2+y^2}$ is the distance from this axis and $\varphi$ is the angle around it.
The reduced field is $\phi(s,z)$, with $s=\sqrt{t^2-\rho^2}$: at a fixed laboratory time $t$, different radii therefore sample different times $s$ in the reduced simulation.
All energy measurements below are made on a surface of constant $t$, not constant $s$.

The collision begins on the axis and spreads to larger transverse radii.
We define the first contact time by $R_g(s_c)=d/2$, and denote the gradient-peak width at contact by $\delta_c$.
The corresponding contact stage, $s=s_c$, reaches radius $\rho$ at laboratory time $t_{\rm hit}(\rho)=\sqrt{s_c^2+\rho^2}$.
We specify the annulus by the interval during which this contact stage sweeps across it: the chosen inner and outer edges are reached at $1.4s_c$ and $1.5s_c$, respectively.
These two factors set the ring's location and thickness. This precise interval was selected from several trial regions in the late-time calibration.
Its radii are therefore $\rho_a=s_c\sqrt{1.4^2-1}$ and $\rho_b=s_c\sqrt{1.5^2-1}$.
We start measuring at time $t_{\rm ref}=1.5s_c$, when the entire region inside the annulus contains collided bubble walls.
The radii remain fixed thereafter, rather than following the expanding collision circle.
The two panels of \cref{fig:gradient-retention-patch} show this same sampling region in laboratory and reduced coordinates.

Although its radial extent is fixed, we enlarge the volume along $z$ to include the outgoing disturbances.
Its axial half-width is
\begin{equation}
    z_{\max}(\rho,t)=4\delta_c+
        \max\!\left(\sqrt{t^2-\rho^2}-s_c,0\right).
    \label{eq:patch-axial-window}
\end{equation}
The sampled volume is thus
\begin{equation}
    \begin{split}
    \mathcal P(t)=\big\{(\rho,\varphi,z):\;&\rho_a\leq\rho\leq\rho_b,\quad
                              0\leq\varphi<2\pi,\\
                           &|z|\leq z_{\max}(\rho,t)\big\},
                           \qquad t\geq t_{\rm ref},
    \end{split}
    \label{eq:patch-region}
\end{equation}
We track the spatial gradient energy in this volume,
\begin{equation}
    E_{\mathcal P}(t)=\frac12\int_{\mathcal P(t)}\mathrm d^3x\,
       [(\partial_z\phi)^2+(\partial_\rho\phi)^2],
    \label{eq:patch-gradient-energy}
\end{equation}
using $\partial_\rho\phi=-(\rho/s)\partial_s\phi$ to reconstruct the transverse gradient from the reduced fields.

\begin{figure}[!ht]
    \centering
    \includegraphics[width=\columnwidth]{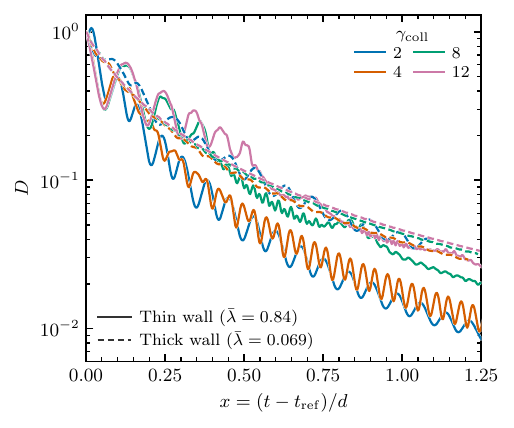}
    \caption{Measured gradient-energy retention $D=E_{\mathcal P}(t)/E_{\mathcal P}(t_{\rm ref})$ in the annular sampling volume of \cref{eq:patch-region}.
    Solid and dashed lines show the thin-wall $\bar\lambda=0.84$ and thick-wall $\bar\lambda=0.069$ cases, respectively, with colours identifying $\gamma_{\rm coll}=2,4,8,12$.}
    \label{fig:gradient-retention-curves}
\end{figure}

\begin{samepage}
We characterize the subsequent relaxation by expressing the gradient energy relative to its value when the measurement begins
\begin{equation}
    D(x;\bar\lambda,\gamma_{\rm coll})
    =\frac{E_{\mathcal P}(t_{\rm ref}+xd)}
           {E_{\mathcal P}(t_{\rm ref})},
    \qquad x=\frac{t-t_{\rm ref}}d.
    \label{eq:absolute-patch-retention}
\end{equation}
Starting from $D(0)=1$, the measured histories in \cref{fig:gradient-retention-curves} show how the gradient energy in the sampling region evolves for different potentials and collision boosts.
\end{samepage}

Our damping prescription describes the gradual loss of a screened contribution rather than its short-time growth and oscillations.
We therefore use a bounded, nonincreasing representation $\widehat D$ of these histories, normalized to $\widehat D(0)=1$, as the retention function entering the kernel in \cref{eq:weight-retention-kernel}.

\begin{figure*}[!ht]
    \centering
    \includegraphics[width=0.87\textwidth]{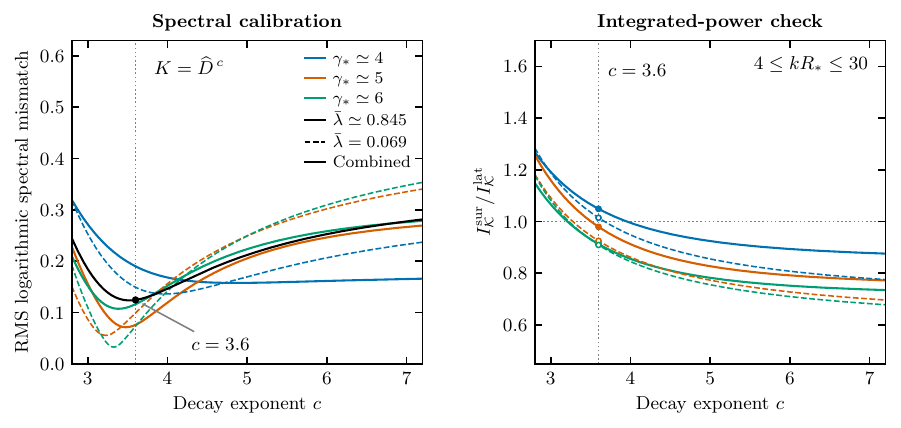}
    \caption{Calibration of the common exponent in $K=\widehat D^{\,c}$ for six many-bubble benchmarks.
    Colours distinguish $\gamma_*\simeq4,5,6$; solid and dashed curves distinguish the thin- and thick-wall potentials, respectively.
    The simulation parameters are listed in \cref{tab:many-bubble-benchmarks}.
    Left: root-mean-square logarithmic spectral residuals over $4\leq kR_*\leq30$, after matching the lattice shell estimator.
    The black curve is $\sqrt{\mathcal L(c)}$ from \cref{eq:decay-calibration-loss}.
    Right: $I_{\mathcal K}^{\rm sur}/I_{\mathcal K}^{\rm lat}$ for the same nominal wavenumber interval.
    This is a separate diagnostic and does not enter the fit.
    The vertical lines and markers identify the adopted exponent $c=3.6$.
    Lattice targets are the pointwise medians of the selected post-screening snapshots.}
    \label{fig:weight-decay-calibration}
\end{figure*}

\subsection{Decay clocks and spectral assembly}

The geometric history tells us when a fraction of a pair's collision circle becomes covered.
These fractions refer to the total circumference, rather than individually tracked material wall elements.
Instead of removing that contribution immediately, we record its covering time and let it decay with the kernel $K_{ij}$ in \cref{eq:weight-retention-kernel}.
For example, if the exposed fraction falls from $0.8$ to $0.6$ at time $u$, with no earlier decaying contribution, the effective weight becomes $0.6+0.2K_{ij}(t-u)$ until the next geometric change.
Each later loss starts its own clock: fractions covered at different times have different decay ages.

The total weight is the exposed fraction plus these remaining contributions,
\begin{equation}
    \mathcal W_{ij}^{\rm eff}(t)
    =\mathcal W_{ij}^{\rm inst}(t)+\sum_a\Delta w_a(t)K_{ij}(t-u_a),
    \label{eq:decaying-geometric-weight}
\end{equation}
The sum includes covering events up to time $t$; $u_a$ is their recorded time and $\Delta w_a(t)$ is the covered fraction still assigned to each event.
For a decreasing geometric weight, $\Delta w_a$ is simply its drop at $u_a$. 
We now simply use this constructed $\mathcal W_{ij}^{\rm eff}(t)$ in \cref{eq:surrogate-physics}.

\begin{figure*}
      \centering
      \includegraphics[width=0.9\textwidth]{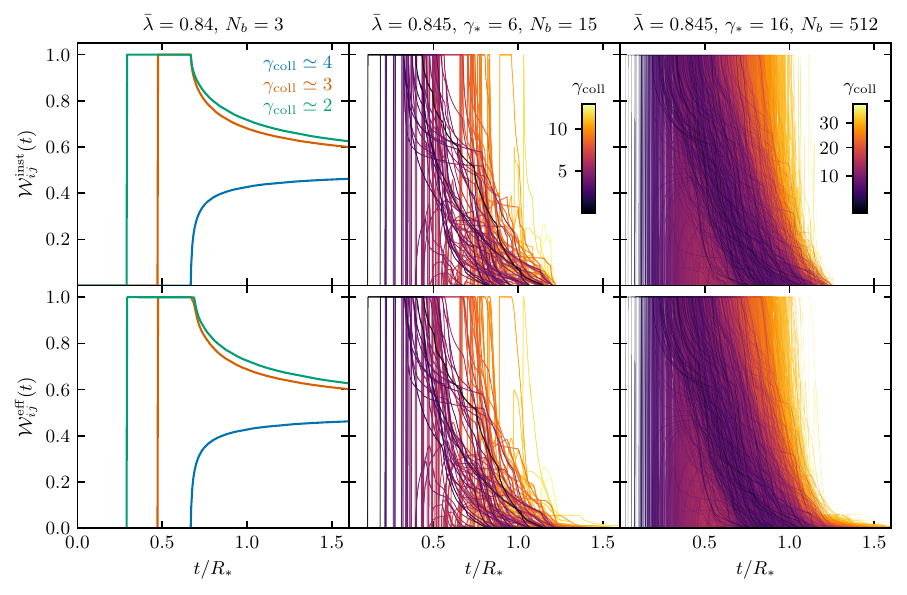}
        \caption{Geometric weights $\mathcal W^{\rm inst}_{ij}(t)$ (top) and $\mathcal W^{\rm eff}_{ij}(t)$ (bottom) of individual bubble collisions. \textit{Left:} the three collisions of the $N_b=3$ simulation shown in \Cref{fig:three-bubble-comparison} at $\bar\lambda=0.84$. \textit{Middle:} all 116 collisions of the $N_b=15$, $\gamma_*=6$ run of \Cref{fig:n-bubble-comparison} at $\bar\lambda=0.845$. \textit{Right:} all 3988 collisions of one realisation with $N_b=512$, $\gamma_*=16$ shown in \Cref{fig:predictions} at $\bar\lambda=0.845$. Line colour indicates $\gamma_{\rm coll}$, with lower $\gamma_{\rm coll}$ drawn on top. Time is in units of the mean pair separation for $N_b=3$ and of $R_*=(L^3/N_b)^{1/3}$ otherwise.}
      \label{fig:weights-example}
\end{figure*}

We illustrate the time-evolution of the weights in \Cref{fig:weights-example} for three different thin-wall scenarios: \textit{(i)} the $N_b=3$ case from \Cref{sec:three-bubble-closure}, \textit{(ii)} the $N_b=15$, $\gamma_*=6$ scenario from \Cref{sec:many-bubble-validation}, and \textit{(iii)} one of the $N_b=512$, $\gamma_*=16$ realisation from \Cref{sec:ultrarelativistic-limit}. The number of individual collisions in these realizations is 3, 116, and 3988, respectively.

For the $N_b=3$ case, all weights are exactly zero at the beginning, as no collisions have taken place yet. At $t(\gamma_{\rm coll}=2)/R_*\approx 0.29$ and $t(\gamma_{\rm coll}=3)/R_*\approx 0.47$, two of the weights suddenly jump to 1, as bubble $i$ starts colliding with bubbles $j$ and $k$. For a short period of time, these two collisions, $ij$ and $ik$, are entirely separate and have no overlap, thus $\mathcal{W}_{ij}=\mathcal{W}_{ik}=1$. At $t(\gamma_{\rm coll}=4)/R_*\approx 0.65$, the third collision, $jk$, would take place, but its initial point of contact happens to lie fully within bubble $i$. The weight $\mathcal{W}_{jk}$ therefore remains zero. As the imaginary collision circle of the $jk$ pair expands, however, part of it eventually leaves bubble $i$ and starts sourcing gravitational waves, resulting in a steep but continuous rise of $\mathcal{W}_{jk}$ from zero at $t/R_*\approx0.67$. At this point, bubble $j$ starts screening the collision circle of $ik$ and bubble $k$ starts screening the collision circle of $ij$, causing the sudden drop in $\mathcal{W}_{ij}$ and $\mathcal{W}_{ik}$.

As our simulation volume was chosen large enough, all three bubbles keep expanding into the false vacuum without completing the phase transition, and thus the weights do not drop to zero within the simulated time. For both $N_b>3$ cases, the picture is qualitatively the same, except that here the box is small enough for bubbles to also collide with periodic images of each other, eventually filling the entire box so that all weights drop to zero.

\subsection{Calibration of the common exponent}
\label{app:decay-calibration}

The isolated-pair histories describe how gradients relax after two walls collide, but do not include the effect of a third wall crossing that region.
We assume that, once the potential and pair-boost dependence is included through $\widehat D$, this additional effect can be described by the same mapping $K=\widehat D^{\,c}$ across the microphysical points studied here.
We calibrate the common exponent against the six many-bubble benchmarks in \cref{tab:many-bubble-benchmarks}, adopting $c=3.6$.
To assess this choice, we vary $c$ while keeping the pair library and geometric histories fixed, and compare the reconstructed late-time spectra with the lattice medians.

The surrogate is binned in the same way as the lattice spectrum and evaluated at the time $T_a$ listed for each configuration in \cref{tab:many-bubble-benchmarks}.
The lattice targets are pointwise medians of the saved spectra after screening has ended, as described in \cref{sec:many-bubble-validation}, rather than snapshots at $T_a$.
For the thick-wall $N_b=48$ benchmark, the surrogate evaluation time is limited by the available pair history.
Writing $\kappa=kR_*$, we quantify the combined disagreement by
\begin{equation}
    \begin{split}
    \mathcal L(c)={}&\frac{1}{6\ln(30/4)}
       \sum_{a=1}^{6}\int_{4}^{30}\frac{\mathrm d\kappa}{\kappa}\\
    &\times\left[\ln\frac{\mathcal P_{{\rm GW},a}^{\rm sur}(\kappa,T_a;c)}
                              {\mathcal P_{{\rm GW},a}^{\rm lat}(\kappa)}\right]^2,
    \end{split}
    \label{eq:decay-calibration-loss}
\end{equation}
where $a$ labels the configurations, each given equal weight.
The normalization averages over the six configurations and the logarithmic width $\ln(30/4)$ of the fitting interval.
The logarithmic ratio measures fractional differences across the peak and falling spectrum, rather than allowing the largest spectral values to dominate the comparison.

Figure~\ref{fig:weight-decay-calibration} shows how both the spectral disagreement and the integrated-energy ratio change with $c$.
We use $c=3.6$ throughout as a common phenomenological prescription, rather than assigning a separately optimized exponent to each configuration.

\section{Implementation of the reduced calculation}
\label[appendix]{app:gpu}
The reduced calculation of \cref{sec:two-bubble-setup} must cover pair boosts up to $\gamma_{ij}=40$ or larger, many cutoff times $T$ per pair, and a grid of frequencies and polar angles. Storing the complete field history $\phi(s,z)$ and evaluating \cref{eq:source-fourier-integral} afterwards would require memory growing approximately as $\gamma_{ij}^4$, exceeding the capacity of a single GPU well below the largest boosts used here. For our largest pair boost, $\gamma_{ij}=40$ at $\bar\lambda=0.845$, the reduced field is evolved on $1.3\times10^6$ grid points in $z$ for $4.3\times10^6$ steps in $s$, of which every fifth enters the source integral. Storing this history in double precision would require about $9\,$TB. We therefore reorganize the calculation as described below, which reduces the memory requirement for this pair to about $12\,$GB of the $40\,$GB available on a single A100 GPU. None of these steps introduces an additional approximation: up to the quadrature of the oscillatory kernels and the order of floating-point summation, the result is identical to integrating the stored history directly.

The field is evolved on the GPU and processed in batches of 128 $s$ slices, each batch retaining the last slice of the previous one so that finite differences across the batch boundary remain available. The source integral $\mathcal T_{ij}(\boldsymbol k;T)$ is linear in the source, so each batch adds its contribution to the real and imaginary parts of each independent component of $\mathcal T_{ij}$, for every frequency, polar angle and cutoff time. Only these linear accumulators persist between batches. The nonlinear step, the tensor contraction and squaring in \cref{eq:reduced-radiation-spectrum}, is applied once after the final batch. The memory requirement is therefore set by the batch size and the output grid rather than by the length of the evolution: for the pair above, the field and derivative buffers of one batch dominate, while the accumulated output for 140 cutoff times, 32 frequencies and 51 polar angles occupies only $11\,$MB.

Since $C_T(t)=1$ for $t<T-t_0/2$, the source integral splits into a part before the window, which is independent of $T$, and the window itself,
  \begin{equation}
      \begin{split}
      \mathcal T_{ij}(\boldsymbol k;T)
      ={}&\int_{t_i}^{T-t_0/2}\mathrm dt\,e^{ikt}S_{ij}(\boldsymbol k,t)\\
      &+\int_{T-t_0/2}^{t_f}\mathrm dt\,C_T(t)\,e^{ikt}S_{ij}(\boldsymbol k,t).
      \end{split}
      \label{eq:plateau-window-split}
  \end{equation}
Here, $t_i$ and $t_f$ are defined in \Cref{eq:source-fourier-integral}.
For increasing cutoff times $T_1<T_2<\dots$, the first term at $T_{n+1}$ is obtained from that at $T_n$ by adding only the interval
  $[T_n-t_0/2,\,T_{n+1}-t_0/2]$.
  The second term has the fixed width set by $t_0$ and is evaluated separately for each $T$.
  The cost of an additional cutoff time is therefore that of its window and the newly added interval, not of a full integration from $t_i$.
  At fixed $s$, the time integrals are evaluated in the variable $t/s$, and the oscillatory kernels from the azimuthal integration are
  integrated with Filon quadrature, with the number of panels set by the number of kernel oscillations in each interval as specified in
  \cref{sec:two-bubble-setup}.

\begin{figure*}[!ht]
      \centering
      \includegraphics[width=0.835\textwidth]{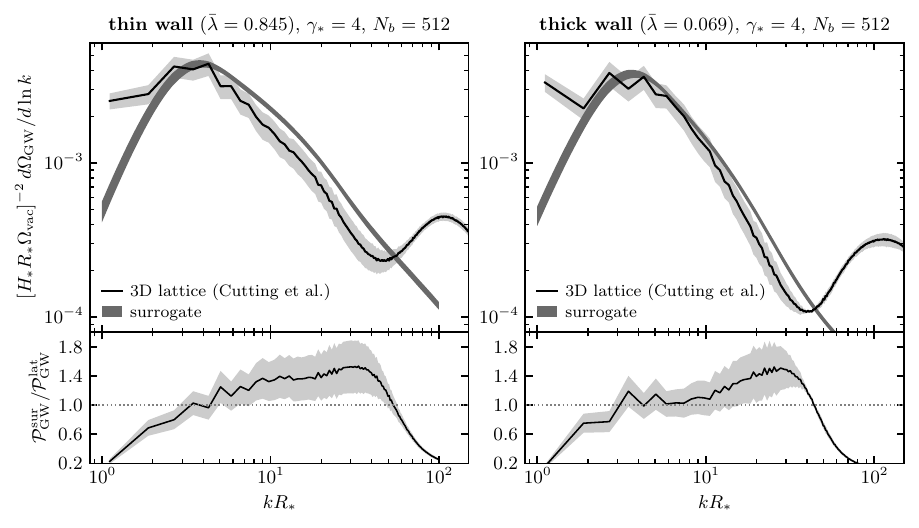}
      \caption{Our surrogate for $N_b=512$ and $\gamma_*=4$ for 16 random bubble placements (envelope), compared to the lattice spectra presented in~\cite{Cutting:2020nla} at $t/R_*=4$ for thin (left) and thick walls (right). Their band is the difference between their high- and medium-resolution runs. In the lower panels, the ratio uses the median over our realizations, and the band combines their resolution difference with the $1\sigma$ scatter of our realizations.}
      \label{fig:literature-comparison}
\end{figure*}

The integrals over $z$ depend on $s$, frequency, and polar angle but not on $T$; they are computed once per batch and reused for all cutoff times. The field derivatives entering the source are evaluated once per $(s,z)$ point and shared by all frequencies and polar angles. For the $\gamma_{ij}=40$ pair, the complete calculation took 15.4 hours on one Nvidia A100 GPU, of which the field evolution takes about two minutes. About $85\%$ of the remainder is spent on the integrals over $z$, which must be repeated for every frequency $k$ and polar angle $\theta$ on each of the $8.6\times10^5$ slices, with oscillatory factors $\cos(kz\cos\theta)$ and $\sin(kz\cos\theta)$. This adds up to about $8\times10^9$ integrals, each over $1.3\times10^6$ grid points. The time integrals, whose kernels oscillate with the Bessel functions from the azimuthal integration and therefore require the Filon quadrature, account for most of the rest.

Because the part of \cref{eq:plateau-window-split} before the window carries the complete history up to $T-t_0/2$, storing it allows a previously computed pair to be extended to later cutoff times without repeating the earlier integration. We use this to extend the time coverage of a library, and to resume runs interrupted by walltime limits.

\section{Comparison with literature}
\label[appendix]{app:literature}
Comparing our reduced calculation and surrogate construction with our own $(3+1)$-dimensional lattice simulations is well suited for validation, as the two computations are largely independent of one another. In addition, we compare with lattice results already published in the literature.

In \Cref{fig:literature-comparison} we compare our surrogate with the gravitational-wave spectra of~\cite{Cutting:2020nla}, obtained from lattice simulations with $N_b=512$ at $\gamma_*=4$ and shown at $t/R_*=4$. Unlike ours, their error band is not based on mode counting but on the difference between a high- and a medium-resolution simulation. As their bubble positions are not available, we evaluate the surrogate for 16 random placements and show the resulting envelope. The surrogate reproduces the peak amplitude and frequency well, but exceeds their spectrum in the UV by up to $\sim40\%$. This is the same
behaviour we find against our own simulations at $\gamma_*\simeq4$ in \Cref{fig:n-bubble-comparison}, whereas the agreement in the UV improves markedly at $\gamma_*\simeq5$--6. The comparison with an independent lattice code therefore confirms that this is a limitation of the surrogate at low $\gamma_*$ rather than a feature of our simulations; we discuss its origin in \cref{sec:many-bubble-validation}.

\end{document}